\documentclass[journal]{IEEEtran}

\usepackage[utf8]{inputenc} 
\usepackage[T1]{fontenc}
\usepackage{color}
\usepackage{float}
\usepackage{amsbsy}
\usepackage{amstext}
\usepackage{setspace}
\usepackage{orcidlink}
\usepackage{graphicx}
\usepackage{bm}
\usepackage{float}
\usepackage{amsmath, amssymb}
\usepackage{gensymb}
\usepackage{upgreek}
\usepackage{breqn}
\usepackage{blindtext}
\usepackage[caption=false]{subfig}
\usepackage{packages/widetext}
\usepackage{packages/dblfloatfix}

\begin{document}

\title{The Azimuthally-Radially Polarized Beam: Helicity and Momentum Densities, Generation and Optimal Chiral Light}

\author{Albert~Herrero-Parareda~\orcidlink{https://orcid.org/0000-0002-8501-5775}
        and~Filippo~Capolino~\orcidlink{https://orcid.org/0000-0003-0758-6182}$^\star$
\thanks{A.~Herrero-Parareda and F.~Capolino are affiliated with the Department of Electrical Engineering and Computer Science, University of California, Irvine,
CA, 92697 USA e-mail: f.capolino@uci.edu.}
\thanks{Manuscript received MONTH DD, YYYY; revised MONTH DD, YYYY.}}

\maketitle
\begin{abstract}
We investigate the optical properties of the azimuthally-radially polarized beam (ARPB), a superposition of an azimuthally polarized beam and a radially polarized beam, which can be tuned to exhibit maximum chirality at a given energy density. This condition is called "optimal chiral light" (OCL) since it represents the maximum local chirality at a given energy density. The transverse fields of an ARPB dominate in the transverse plane but vanish on the beam axis, where the magnetic and electric fields are purely longitudinal. This spatial separation between transverse and longitudinal components leads to vanishing linear and angular momentum densities on the axis, where only the energy and helicity densities associated with the longitudinal fields persist. The ARPB does not have a phase variation around the beam axis and nonetheless exhibits a power flow around the beam axis that causes a longitudinal orbital momentum density. We introduce a concise notation for the ARPB and provide field quantities, especially for the optimally chiral configuration. The ARPB shows promise for precise one-dimensional chirality probing and enantioseparation of chiral particles along the beam axis. We propose a setup to generate ARPBs with controlled chirality and orbital angular momentum.
\end{abstract}

\section{Introduction}
\label{ch:intro}

Chiral molecules, also known as enantiomers, are not superimposable on their mirror images \cite{riehl_mirror_2011, barron_true_2013}. They have the same constitution but follow different chemical processes \cite{ribeiro_chiral_2018}. When interacting with chiral light, which also has non-superimposable mirror images, enantiomers also have different optical properties \cite{tang_optical_2010, bliokh_characterizing_2011}. In 2016 it was reported that $50\%$ of FDA-approved drugs have chiral active ingredients \cite{Mane_racemic_2016}, where one enantiomer initiates the desired chemical reaction and the other one acts as an antagonist \cite{carvalho_review_2006}. Therefore, it is of supreme importance to develop efficient techniques to probe the chirality of enantiomers \cite{jia_probing_2011, poulikakos_optical_2016, zhao_nanoscopic_2017} and to separate them \cite{li_enantioselective_2019, solomon_nanophotonic_2020}. Ref.~\cite{bradshaw_chirality_2015} reports that a system combining chirality and optical trapping could be used to separate enantiomers. The chiroptical response of natural chiral systems, however, is very weak \cite{hentschel_three_2012, du_chiral_2020}. These responses depend on the chirality of light, as observed theoretically in Refs.~\cite{tkachenko_helicity_2014, hayat_lateral_2015, biagioni_chiral_2019}, and experimentally in Ref.~\cite{kravets_optical_2019}. The optical chirality is quantified by the helicity density \cite{cameron_optical_2012}, and the concepts of helicity maximization and optimally chiral light (OCL) were introduced in Ref.~\cite{hanifeh_optimally_2020}. A monochromatic field propagating in free space is optimally chiral (OC) when its helicity density $h$ reaches its upper bound, $|h|= u/\omega$, where $u$ is the energy density of light and $\omega$ its angular frequency. When light is not optimal chiral, one has $|h| < u/\omega$.  
As shown in Ref.~\cite{hanifeh_optimally_2020}, circularly polarized light (CPL) is an example of OCL, and that is why CPL has been used so successfully to detect the chirality of matter. However, the results in Ref.~\cite{hanifeh_optimally_2020} show that there are countless other examples of structured light that are OCL, and the special beam studied in this paper is a useful and simple example. 

Emerging strategies attempt to maximize chiral light-matter interactions by enhancing the optical chirality using dielectric \cite{solomon_enantiospecific_2019, mohammadi_accessible_2019, solomon_nanophotonic_2020, li_tunable_2021} or plasmonic nanostructures \cite{martin_tailoring_2012, valev_chirality_2013, luo_plasmonic_2017}. It is shown in \cite{biagioni_quasistatic_2015} that the average chirality of the near field of plasmonic nanostructures cannot be significantly higher than that in freely-propagating plane waves. These studies so far have only considered nanostructures illuminated with CPL, which only displays optical chirality (i.e., helicity) in the transverse plane perpendicular to the beam propagation direction, or with linearly polarized light \cite{beutel_enhancing_2021} for enhanced optical rotation measurements. However, Gúzman Rosales et al. \cite{rosales_light_2012} highlighted the importance of longitudinal (i.e., along the beam propagation direction) field components in chiral interactions, and Kamandi et al. \cite{kamandi_unscrambling_2018} showed that beams with only longitudinal chirality are useful for unambiguous chirality probing. In this context, we define transverse and longitudinal optical chirality as the contributions associated solely with the transverse and longitudinal field components, respectively, as in Refs.~\cite{kamandi_unscrambling_2018, ye_enhancing_2021, sifat_force_2022}. A beam with transverse chirality can only reveal the transverse chirality of the sample, which could be mistaken with its anisotropy as the sample is probed in two directions ($x$ and $y$) \cite{kamandi_unscrambling_2018, lininger_chirality_2022}. In contrast, a beam with just longitudinal chirality only probes the sample along one direction ($z$).

\begin{figure}[h]
    \centering
    \includegraphics[width = 0.4\textwidth]{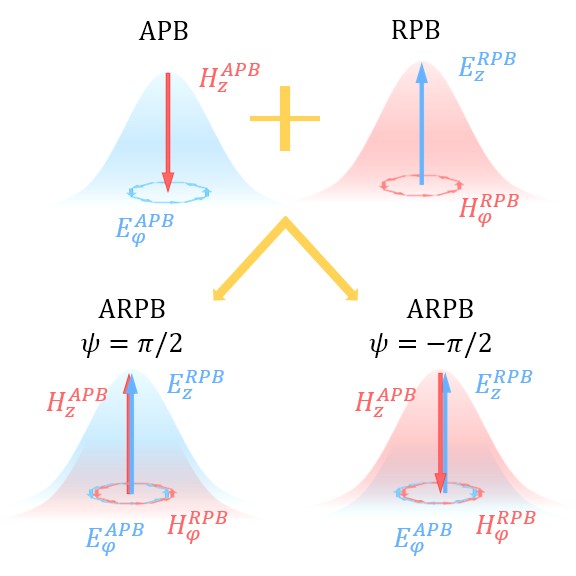}
    \caption{Conceptual representation of the APB, the RPB, and the ARPB as their combination with a phase delay of $\psi = \pm \pi/2$. The electric field is represented in blue and the magnetic field in red. The radial components of the fields are not shown. The bell curve denotes the Gaussian-like parameter $f$. In the ARPB with $\psi = \pi/2$, the electric field $\mathbf{E}$ (blue bell curve) leads in time to the magnetic field $\mathbf{H}$ (red bell curve), and the opposite (lagging) occurs when $\psi = -\pi/2$.}
    \label{fig:BeamRepresentation}
\end{figure}

\begin{figure}[t]%
\centering
\subfloat[]{\includegraphics[width=\linewidth]{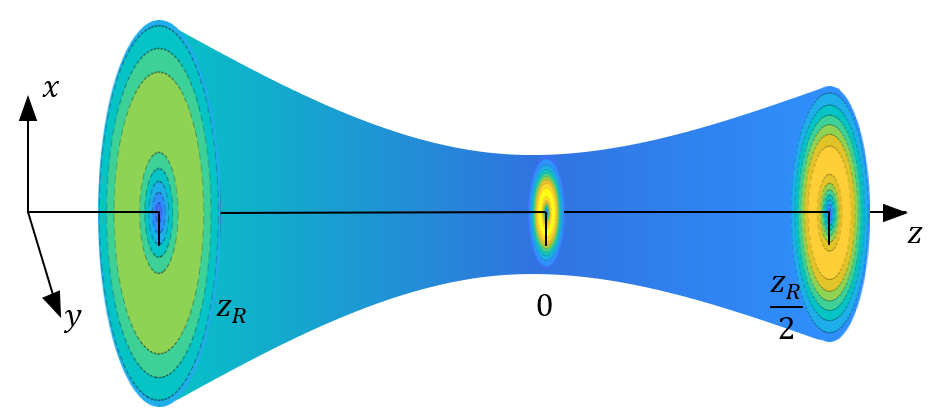}}
\hfill
\subfloat[]{\includegraphics[width=\linewidth]{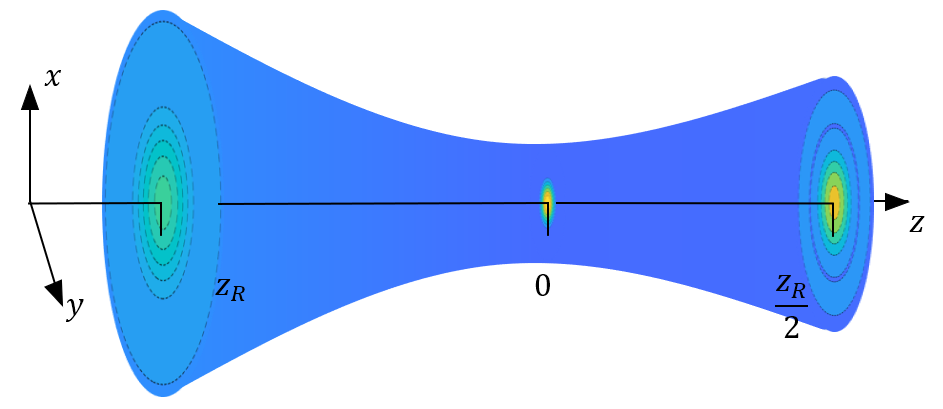}}

\caption[\linewidth]{Schematic representation of propagation and localization of the (a) transverse $\mathbf{E}_\perp = E_\rho \hat{\bm{\rho}}+ E_\varphi \hat{\bm{\varphi}}$, and (b) longitudinal $E_z$ fields of an ARPB with $w_0=\lambda=400$ nm. The $E_z$ field is localized on the axis. The same happens for the magnetic field, with $H_z$ localized on the axis.}
\label{fig:BeamProp}
\end{figure}

A beam with purely longitudinal chirality on the optical axis is the azimuthally-radially polarized beam (ARPB) \cite{kamandi_unscrambling_2018}, a superposition of an azimuthally polarized beam (APB) \cite{veysi_focused_2016, guclu_photoinduced_2016, zeng_exclusive_2018, zeng_sharply_2018, blanco_ultraintense_2019} and a radially polarized beam (RPB) \cite{zhan_cylindrical_2009, rui_trapping_2014, lin_cladding_2014}. In Ref.~\cite{eisman_exciting_2018}, the authors tightly focus an achiral beam resulting from the in-phase superposition of an APB and an RPB to excite a chiral dipole moment in an achiral nanostructure. Here, instead, we examine out-of-phase superpositions of these two beams, which results in a chiral ARPB. The phase-shift $\psi$ and the ratio of amplitudes $\hat{V}$ between the APB and the RPB can be tuned to produce an ARPB with optimal chirality in all directions, which we refer to as an optimally chiral ARPB (OC-ARPB). This occurs when $\psi = \pm \pi/2$ and $\hat{V}=1$ \cite{hanifeh_optimally_2020}. Figure~\ref{fig:BeamRepresentation} displays schematic views of an APB, an RPB, and the two configurations of an OC-ARPB. For $0 < \psi < \pi$, the electric field leads the magnetic field, while the opposite occurs for $-\pi< \psi < 0$. The transverse fields of the ARPB vanish on the beam axis, where the longitudinal fields persist \cite{kamandi_unscrambling_2018, hanifeh_optimally_2020, hanifeh_helicity_max_2020, jiang_theory_2021}. This spatial separation is shown in Figures~\ref{fig:BeamProp}(a) and~(b), which respectively depict the propagation of the magnitude of the transverse $\mathbf{E}_\perp = E_\rho \hat{\bm{\rho}}+ E_\varphi \hat{\bm{\varphi}}$ and longitudinal $E_z$ electric fields of the ARPB. The ARPB only displays longitudinal chirality on the beam axis and can be used to confidently probe the chirality of an anisotropic sample (as opposed to CPL). Moreover, preliminary theoretical results show that illuminating dielectric nanospheres with an OC-ARPB results in almost twice the helicity density enhancement than using a CP Gaussian beam \cite{hanifeh_optimally_2020}. Other investigations into beams with longitudinal chirality, including Gúzman Rosales et al \cite{rosales_light_2012}, study the optical chirality of beams with orbital angular momentum (OAM) \cite{rosales_light_2012, koksal_optical_2022, koksal_hopf_2022, babiker_zero_2022, wozniak_interaction_2019, forbes_optical_2022, forbes_enantioselective_2022}.

In this paper, we further advance the concept of a chiral ARPB, defined in terms of an arbitrary relative amplitude $\hat{V}$ and phase shift $\psi$ between the APB and the RPB, as in Refs.~\cite{hanifeh_optimally_2020, hanifeh_helicity_max_2020, hanifeh_helicity_max_planar_2020, jiang_theory_2021}. We introduce a concise notation for the APB and the RPB in Section~\ref{ch:CylPolBasis}. The ARPB is described in Section~\ref{ch:ARPB}, and the analytical expressions for its field quantities are provided in Section~\ref{ch:FieldQuantities}. In Section~\ref{ch:OptimallyChiral}, we focus on the OC-ARPB and its exceptional electric-magnetic symmetry. To support future experimental studies, we discuss setups that generate an ARPB in Section~\ref{ch:GenerationManipulation}. The conclusions are found in Section~\ref{ch:Conclusion}. 

\section{Cylindrical polarization basis}
\label{ch:CylPolBasis}

The non-zero field components of the APB in the cylindrical basis ($\rho$, $\varphi$, $z$) are

\begin{equation}
    \begin{array}{c}
      E_{\varphi}^{\mathrm{APB}}=\frac{\rho}{w^2}V_Af, \\
     H_{\rho}^{\mathrm{APB}}=-\frac{\rho}{w^2\eta_0}V_Af\left(A_\rho + iB_\rho\right),\\
     H_{z}^{\mathrm{APB}}=-\frac{2i}{kw^2\eta_0}V_Af\left(A_z + iB_z\right),
    \end{array}
    \label{eq:APBDef}
\end{equation}

and those of the RPB
    
\begin{equation}
    \begin{array}{c}
    H_{\varphi}^{\mathrm{RPB}}=\frac{\rho}{w^2\eta_0}V_Rf, \\
    E_{\rho}^{\mathrm{RPB}}= \frac{\rho}{w^2}V_Rf\left(A_\rho + iB_\rho \right),\\
    E_{z}^{\mathrm{RPB}}=\frac{2i}{kw^2}V_Rf\left(A_z + iB_z\right),
    \end{array}
    \label{eq:RPBDef}
\end{equation}

where $V_A$, $V_R$ are their complex amplitudes (with unit of Volts), respectively. The dimensionless shorthand parameters are

\begin{equation}
\begin{array}{c}
    f = \frac{2}{\sqrt\pi}e^{-(\rho/w)^{2}\zeta}e^{-2i\tan^{-1}(z/z_{\text{R}})}e^{ikz}, \\
    A_\rho = 1+\frac{1}{kz_{R}}\frac{\rho^{2}-2w_{0}^{2}}{w^{2}} +\left(\frac{2z\rho}{w^2 k z_R }\right)^2, \\
    B_\rho = -\frac{4}{w^{2}}\frac{1}{k^{2}}\frac{z}{z_{R}}\left(1-\frac{\rho^{2}}{w^{2}}\right), \\
    A_z = 1-\frac{\rho^2}{w^2}, \\
    B_z = \frac{z}{z_R}\frac{\rho^2}{w^2},
\end{array}
\label{eq:Simplification}
\end{equation}

where $w=w_0\sqrt{1+(z/z_R)^2}$ represents the beam radius, and $w_0$ is defined as half the beam waist parameter at $z=0$. The Gouy phase is given by $\zeta = 1-iz/z_R$, and the Rayleigh range is expressed as $z_R = \pi w_0^2/\lambda$, where $\lambda$ is the free-space wavelength. The wavenumber is $k=2\pi/\lambda$. The characteristic impedance of free space is given by $\eta_0 = \sqrt{\mu_0 / \varepsilon_0}$, where $\mu_0$ and $\varepsilon_0$ are the absolute permeability and permittivity in vacuum, respectively. In this paper, we consider monochromatic fields with the implicit time dependence $e^{-i\omega t}$, where $\omega$ is the angular frequency of the fields and $c$ is the speed of light in vacuum.

The dimensionless parameters $A_\rho$, $B_\rho$, $A_z$, and $B_z$ described in Equation~(\ref{eq:Simplification}) are real, but the parameter $f$ is complex and related to the normalized Laguerre-Gaussian (LG) modes \cite{allen_orbital_1992, veysi_focused_2016} 

\begin{equation}
    u_{\pm 1,0}=\frac{2\rho}{\sqrt{\pi}w^2}e^{-\left(\rho/w\right)^2\zeta}e^{-2i\tan^{-1}\left(z/z_R\right)}e^{\pm i\varphi},
\label{eq:LGMode}
\end{equation}

as $f=\frac{w^2}{\rho}u_{\pm1,0}e^{\mp i\varphi}e^{ikz}$. Further elaboration on this subject can be found in Section A of the Supporting Information, where we show the APB electric field in different polarization bases. Comparing Equations~(\ref{eq:APBDef}) and~(\ref{eq:RPBDef}) shows that the APB and the RPB are electromagnetically dual. When they have the same beam parameters, their fields are related as \cite{jiang_theory_2021}

\begin{equation}
    \begin{array}{c}
        \mathbf{E}^{\text{APB}}=\eta_{0}\mathbf{H}^{\text{RPB}}\left(\frac{V_{\text{A}}}{V_{\text{R}}}\right),\\
        \mathbf{E}^{\text{RPB}}=-\eta_{0}\mathbf{H}^{\text{APB}}\left(\frac{V_{\text{R}}}{V_{\text{A}}}\right).
    \end{array}
    \label{eq:EMDuality}
\end{equation}

The APB/RPB only has non-zero longitudinal magnetic/electric field on the axis ($\rho = 0$).

\section{The azimuthally-radially polarized beam}
\label{ch:ARPB}

The ARPB is formed by adding an APB and an RPB, with a complex amplitude ratio given by $V_A / V_R = \hat{V} e^{i\psi}$. The parameter $\hat{V}$ monitors the relative magnitudes of the two beams, while their relative phase shift $\psi$ is referred to as the phase parameter as in Ref.~\cite{kamandi_unscrambling_2018}. The combined fields of the ARPB are

\begin{widetext}
\begin{equation}
\begin{array}{c}
        \mathbf{E}^{\text{ARPB}}=\mathbf{E}^{\text{APB}}+\mathbf{E}^{\text{RPB}} = \frac{fV_R}{kw^2}\left[k\rho \left(A_\rho + iB_\rho\right)\,\hat{\bm{\rho}} + k\rho \hat{V}e^{i\psi}\,\hat{\bm{\varphi}}+2i\left(A_z + iB_z\right)\,\hat{\bm{z}}\right], \\
        \mathbf{H}^{\text{ARPB}}=\mathbf{H}^{\text{APB}}+\mathbf{H}^{\text{RPB}} = - \frac{fV_R}{kw^2\eta_0}\left[k\rho \hat{V}e^{i\psi}\left(A_\rho + iB_\rho\right)\,\hat{\bm{\rho}} - k\rho\,\hat{\bm{\varphi}}+2i\hat{V}e^{i\psi}\left(A_z + iB_z\right)\,\hat{\bm{z}}\right],
\end{array}
\label{eq:ARPBDef}
\end{equation}
\end{widetext}

which are purely longitudinal on the beam axis. Note that the field components in cylindrical coordinates do not show an azimuthal dependence. In Ref.~\cite{veysi_focused_2016}, the authors demonstrate that as $w_0$ of an APB decreases, its longitudinal magnetic field component $H_z$ increases at a faster rate compared to its transverse fields $\mathbf{E}_\perp$ and $\mathbf{H}_\perp$. As a result of the APB/RPB duality shown in Eq.~(\ref{eq:EMDuality}), this phenomenon is also observed in the longitudinal electric field component $E_z^{RPB}$ of an RPB. Consequently, the longitudinal fields of the ARPB increase more rapidly than the transverse fields as $w_0$ decreases.

\section{ARPB field quantities}
\label{ch:FieldQuantities}

We provide the cycle-averaged densities of specific field quantities of the ARPB that play a critical role in local chiral light-matter interactions \cite{tang_optical_2010, hayat_lateral_2015}. The cycle-averaged energy density of a monochromatic EM field is $u = \varepsilon_0 |\mathbf{E}|^2/ 4 + \mu_0|\mathbf{H}|^2/4$ \cite{angelsky_structured_2020}, and the complex Poynting vector is $\mathbf{S}=\left(\mathbf{E}\times\mathbf{H}^*\right) / 2$ \cite{bliokh_extraordinary_2014}. Its real part is associated with the linear momentum density $\mathbf{p}=\Re(\mathbf{S})/c^2$ as in \cite{bliokh_extraordinary_2014}, while the imaginary part is associated with the reactive energy flux density of the field. The spin angular momentum (SAM) density is $\bm{\sigma} = -\frac{\varepsilon_0}{4i\omega }\left(\mathbf{E}\times\mathbf{E}^*\right)-\frac{\mu_0}{4i\omega }\left(\mathbf{H}\times\mathbf{H}^*\right)$, and its helicity density is $h = \frac{1}{2\omega c}\Im\left(\mathbf{E}\cdot\mathbf{H}^*\right)$ \cite{hanifeh_optimally_2020}. While the energy and SAM densities naturally split into electric and magnetic field contributions, i.e., $u = u_e + u_m$ and $\bm{\sigma} = \bm{\sigma}_e + \bm{\sigma}_m $, the Poynting vector $\mathbf{S}$ and the helicity density $h$ depend on the interaction between the electric and magnetic fields of the beam. In this section, our focus is exclusively on the final expressions for the field quantities of the ARPB. Additional information about the field quantities of the APB, the RPB, and the ARPB can be found in Section B of the Supporting Information. To make the field quantities easier to comprehend, we express them as a combination of a normalization constant (denoted by the subscript $0$) and a dimensionless term. The energy density of an ARPB is

\begin{equation}
  \begin{array}{c}
  u^{\mathrm{ARPB}} = \frac{1}{2}u_0\left(1 + \hat{V}^2\right)\left[(k\rho)^2\left(1 + A_\rho^2 + B_\rho^2\right) + 4\left(A_z^2 + B_z^2\right)\right],
  \end{array}
  \label{eq:ARPBEnergyDensity}
\end{equation}

where $u_0 = \frac{\varepsilon_0|f|^2}{2k^2w^4}|V_R|^2$. The energy density of an ARPB describes an annular ring in the transverse plane. The components of the Poynting vector $\mathbf{S}^{\mathrm{ARPB}}$ of an ARPB are 

\begin{equation}
    \begin{array}{c}
         S_\rho^{\mathrm{ARPB}} = iS_0\left[\right. \hat{V}^2(A_z -iB_z)  - (A_z + iB_z)\left.\right],\\
         S_\varphi^{\mathrm{ARPB}} = -2S_0\hat{V}\left(\sin\psi+i\cos\psi\right)\left(A_\rho A_z + B_\rho B_z\right),\\
         S_z^{\mathrm{ARPB}} = \frac{1}{2}S_0 k \rho\left[\right.\hat{V}^2(A_\rho -iB_\rho) + (A_\rho +iB_\rho)\left.\right].
    \end{array}
    \label{eq:ARPBPoyntingVector}
\end{equation}

where $S_0 = \frac{\rho|f|^2}{k\eta_0w^4}|V_R|^2= 2\rho \omega u_0$. Note that $S_0$ vanishes on the beam axis, while $u_0$ does not. When $\hat{V}=1$, the longitudinal and radial energy flux density, $S_z^{\mathrm{ARPB}}$ and $S_{\rho}^{\mathrm{ARPB}}$, are purely real. Moreover, near its focal plane ($z \approx 0$), the radial component of the Poynting vector of an ARPB with $\hat{V}=1$ is negligible: $S_\rho^{\mathrm{ARPB}}\propto B_z \approx 0$ $\text{W/m}^2$ (see Eq.~(\ref{eq:Simplification})). Therefore, most of the energy density flows in the $z$ direction and the ratio of active versus reactive energy flux density in the azimuthal direction depends on the phase parameter $\psi$. For $\psi \neq 0,\pi$, which is the case for chiral ARPBs, there is a nonzero azimuthal linear momentum density $p_\varphi = \Re(S_\varphi)/c^2$. While the ARPB is not a conventional vortex beam since it does not have a phase variation around the axis (usually characterized by $e^{il\varphi}$) \cite{leach_vortex_2005}, it exhibits a power flow around the axis that can be controlled by $\psi$. Therefore, it necessarily implies the existence of an orbital angular momentum (OAM) density in the longitudinal direction \cite{allen_poynting_2000, speirits_waves_2013, barnett_natures_2016}. The longitudinal OAM density $l_z^{\mathrm{ARPB}}$ of the ARPB arises from the propagation of energy in the azimuthal direction $p_\varphi^{\mathrm{ARPB}}$, which is, in essence, a rotation of the beam around the axis. This rotation appears as a result of the out-of-phase combination of an APB and an RPB, which do not individually exhibit an azimuthal power flow (see Eq.~(S5) of the Supplementary Information, Section B, for further details). The longitudinal OAM density of a field in cylindrical coordinates is calculated as $l_z = p_\varphi \rho -\sigma_z$ (see Eq.~(S8) and (S9) of the Supplementary Information, Section B, for more information), which for an ARPB is

\begin{equation}
    l_z^{\mathrm{ARPB}} =l_0 \hat{V}\sin(\psi)\left[2(A_\rho A_z + B_\rho B_z) + A_\rho\right],
    \label{eq:OAMARPB}
\end{equation}

where $l_0 = - \frac{\varepsilon_0\rho^2|f|^2}{\omega w^4}|V_R|^2 $. The longitudinal OAM density vanishes on the beam axis, where the fields are purely longitudinal. Remarkably, the OAM density-carrying ARPB is generated from a phase-shifted superposition of two beams with zero longitudinal OAM densities, the APB and the RPB. A chiral ARPB carries an $l_z$ that reaches a maximum for $\psi=\pm\pi/2$, which is the case for OCL as discussed in Section~\ref{ch:OptimallyChiral}. Additionally, the OAM density of an ARPB can be continuously tuned by varying the phase parameter $\psi$, in contrast with the discrete azimuthal phase dependence $e^{il\varphi}$ of LG vortex beams, where $l$ is the topological charge. As shown later in Section~\ref{ch:GenerationManipulation}, the value of $\psi$ is readily accessible by means of standard optical elements (beam splitters, mirrors, and wave plates), while varying $l$ generally requires elaborate optical elements, including Q-plates \cite{rubano_Q-plate_2019}, spatial light modulators (SLMs)\cite{forbes_creation_2016}, spiral phase plates \cite{schemmel_modular_2014}, or cylindrical lenses \cite{beijersbergen_astigmatic_1993, padgett_orbital_2002}.

The components of the SAM density $\bm{\sigma}^{\mathrm{ARPB}}$ of the ARPB are

\begin{equation}
    \begin{array}{c}
    \sigma_\rho^{\mathrm{ARPB}} = 2\sigma_0 \hat{V}\sin(\psi)B_z, \\
    \sigma_\varphi^{\mathrm{ARPB}} = -\sigma_0\left(1 + \hat{V}^2\right)\left[A_\rho A_z + B_\rho B_z\right], \\
    \sigma_z^{\mathrm{ARPB}} = \sigma_0 k\rho \hat{V}\sin(\psi)A_\rho,
    \end{array}
    \label{eq:ARPBSpin}
\end{equation}

where $\sigma_0 = \frac{\varepsilon_0\rho|f|^2}{\omega k w^4}|V_R|^2=S_0/(\omega c)$ vanishes on axis. While the azimuthal SAM density $\sigma_\varphi^{\mathrm{ARPB}}$ is the sum of the SAM densities of the individual APB and RPB (see Supporting Information, Section B), the radial and longitudinal components of the SAM density arise from the interaction between the two beams and therefore depend on the phase parameter $\psi$.

The cycle-averaged helicity density $h$ is naturally split between the contributions from each field component, i.e., $h = h_\rho + h_\varphi + h_z$, and also between the transverse $h_\perp = h_\rho + h_\varphi$ and longitudinal $h_z$ helicity densities. The components of the helicity density $h^{\mathrm{ARPB}}$ of an APRB are

\begin{equation}
    \begin{array}{c}
         h_\rho = h_0\hat{V}\sin(\psi)(k\rho)^2\left(A_\rho^2 + B_\rho^2\right),\\
         h_\varphi = h_0\hat{V}\sin(\psi)(k\rho)^2,\\
         h_z = 4h_0\hat{V}\sin(\psi)\left(A_z^2 + B_z^2\right),
    \end{array}
    \label{eq:HelicityDensityComponents}
\end{equation}

where $h_0 = \frac{\varepsilon_0|f|^2}{2\omega k^2w^4}|V_R|^2 $. The normalization parameters of the energy $u_0$ and helicity $h_0$ densities are related as $h_0=u_0/\omega$. Note that $h_0$ and $u_0$ do not vanish on the axis, while $S_0$, $\sigma_0$, and $l_0$ do vanish there, and that is why there is a special relation between $h_0$ and $u_0$. The total helicity density of an ARPB is also related to its energy density (shown in Eq.~(\ref{eq:ARPBEnergyDensity})) as

\begin{equation}
    h^{\mathrm{ARPB}}=\frac{u^{\mathrm{ARPB}}}{\omega}\frac{2\hat{V}}{1+\hat{V}^2}\sin(\psi).
    \label{eq:HelicityEnergy}
\end{equation}

For $\hat{V} = 1$, the helicity density is in agreement with the expressions in Eq.~(13.28) and~(13.29) of Ref.~\cite{jiang_theory_2021}. For $\psi=0, \pi$, the interaction between the fields of the APB and the RPB is minimized; the energy $u$, linear momentum $\mathbf{p}$, SAM $\bm{\sigma}$, longitudinal OAM $l_z$, and helicity $h$ densities of the resulting ARPB are the sum of those of the APB and the RPB (see Section B of the Supporting Information for details). Conversely, for $\psi = \pm \pi/2$, which corresponds to the OC-ARPB as shown in Section~\ref{ch:OptimallyChiral}, the interaction between the APB and the RPB is maximized. The OC-APRB displays maximum linear momentum, SAM, OAM, and helicity densities. In essence, the phase parameter $\psi$ mediates the interaction between the fields of the APB and the RPB. While the longitudinal OAM $l_z$ and helicity densities are proportional, the latter is not a by-product of the former. In fact, both field quantities originate from the out-of-phase superposition of the APB and the RPB, which is determined by the phase parameter $\psi$, in contrast to what occurs for focused LG beams $u_{l,p}$ \cite{forbes_optical_2022}, where the helicity density is proportional to the topological charge $l$.

Moreover, the ARPB displays an exceptional spatial separation between the longitudinal and transverse fields, which vanish on the beam axis. As a consequence, the energy $u$, linear momentum $\mathbf{p}$, SAM $\bm{\sigma}$, longitudinal OAM $l_z$, and helicity $h$ densities describe an annular ring in the transverse plane. On the $z$ axis, only the contribution of the longitudinal fields to the energy and helicity densities, shown respectively in Eqs.~(\ref{eq:ARPBEnergyDensity}) and~(\ref{eq:HelicityEnergy}), persist. 

\begin{figure*}[t]%
\centering
\subfloat[]{\includegraphics[width=0.32\textwidth]{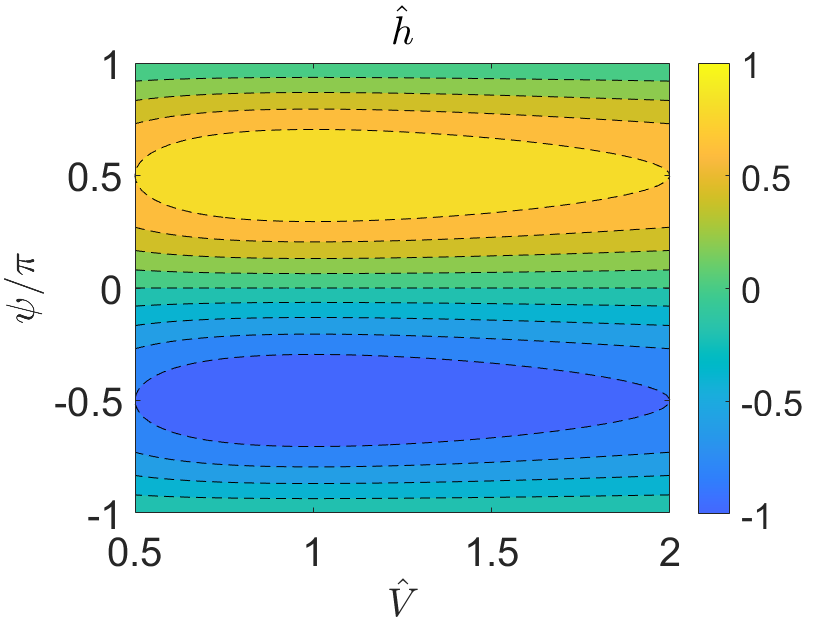}}
\hfill
\subfloat[]{\includegraphics[width=0.32\textwidth]{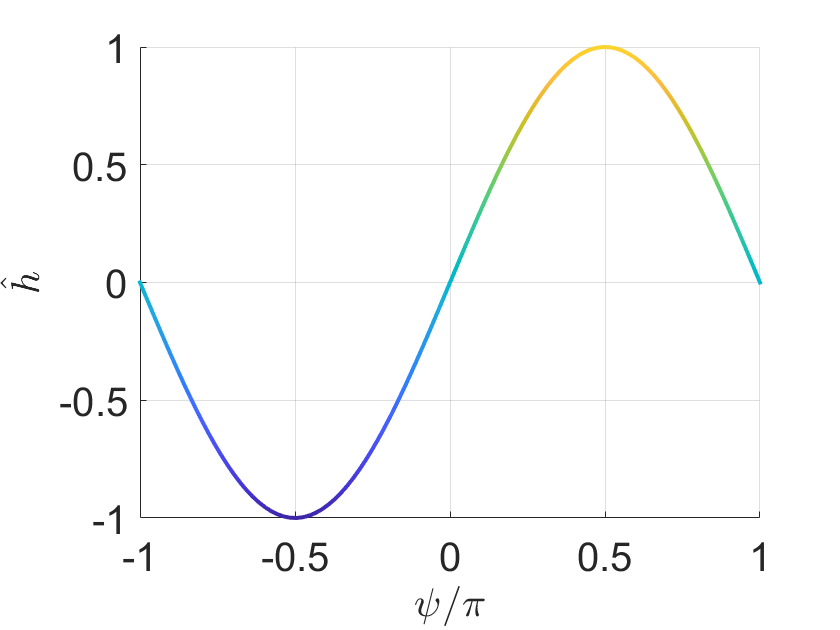}}
\hfill
\subfloat[]{\includegraphics[width=0.32\textwidth]{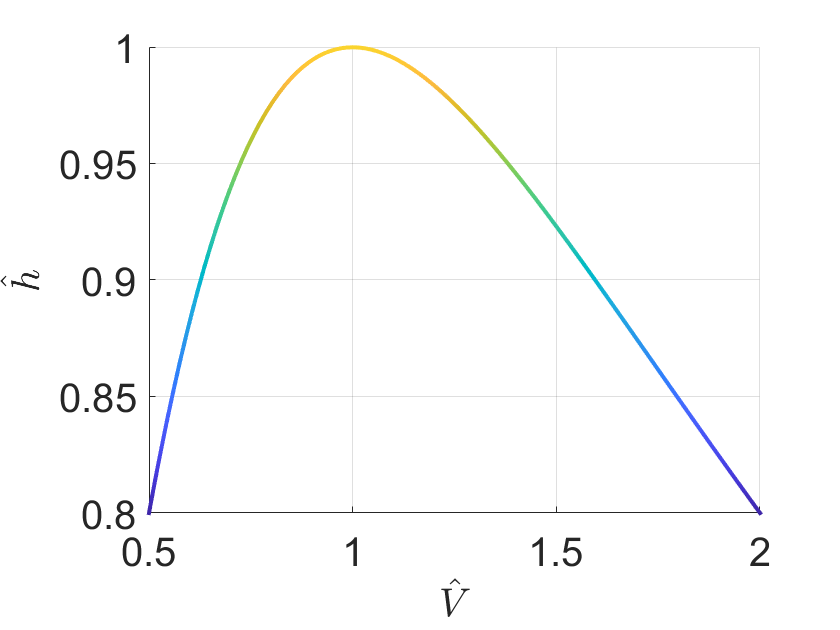}}
\caption[\linewidth]{Normalized helicity density $\hat{h}$ of an ARPB in terms of $\hat{V}$ and $\psi$, calculated as $\hat{h}=hu/\omega$. (a) Colormap of $\hat{h}$ for $\hat{V} \in [0,2]$ and $\psi \in [-\pi, \pi]$. The maximum magnitude for the normalized helicity density occurs for OC-ARPBs that is $V_A = \pm i V_R$. (b) The normalized helicity density for $\hat{V}=1$ and $\psi \in [-\pi, \pi]$. The maximum values are found for $\psi = \pm\pi/2$. (c) The normalized helicity density for $\psi=\pi/2$ and $\hat{V} \in [0,2]$. The magnitude of the normalized helicity density reaches unity ($\hat{h}=1$) for $\hat{V}=1$ and $\psi=\pm \pi/2$, i.e., $V_A=\pm iV_R$.}
\label{fig:Helicity}
\end{figure*}

\section{Optimally chiral ARPB}
\label{ch:OptimallyChiral}

It was demonstrated in Ref.~\cite{hanifeh_optimally_2020} that the helicity density of a chiral field is locally bounded by its energy density as

\begin{equation}
    -\frac{u}{\omega} \leq h \leq \frac{u}{\omega}.
    \label{eq:HelicityBound}
\end{equation}

By definition, OCL is a local descriptor that structured light reaches the upper or lower bound of the helicity density. It occurs for fields whose phasors locally satisfy $\mathbf{E}=\pm i \eta_0\mathbf{H}$, which is referred to as the optimal chirality condition \cite{hanifeh_optimally_2020}. This is equivalent to the condition that the field phasors of OCL must satisfy

\begin{equation}
    \begin{array}{c}
         \nabla\times\mathbf{E}=\pm k\mathbf{E},  \\
         \nabla\times\mathbf{H}=\pm k\mathbf{H}.
         \label{eq:NablaTimesEisE}
    \end{array}
\end{equation}

Thus, the optical chirality density $C=(1/4) (\varepsilon_0 \mathbf{E}^*\cdot\nabla\times \mathbf{E} + \mu_0\mathbf{H}^*\cdot\nabla\times\mathbf{H})$ \cite{mackinnon_on_2019}, which evaluates the collinearity between the fields and their curls, is also maximum for OCL. For time-harmonic fields in free space, the optical chirality and the helicity densities are proportional, with $C=\omega^2 c^{-1} h$ \cite{hanifeh_optimally_2020}. As a result, the magnitude of optical chirality has an upper bound given by the energy density,  $|C|\leq ku$, which is reached for OCL.

In this paper, we find convenient to use the normalized (and dimensionless) helicity density $\hat{h}=h \omega/u$, whose magnitude is bounded by unity, i.e., $-1\leq \hat{h} \leq 1$. It is naturally split into its transverse $\hat{h}_\perp = h_\perp \omega/u$ and longitudinal $\hat{h}_z = h_z \omega/u$ components (with $|\hat{h}_\perp+\hat{h}_z|\leq 1$). The normalized helicity density $\hat{h}^{\mathrm{ARPB}}$ of the ARPB is reduced to

\begin{equation}
    \hat{h}^{\mathrm{ARPB}} =\frac{2\hat{V}}{1+\hat{V}^2}\text{sin}(\psi),
    \label{eq:ARPBNormalizedHelicityDensity}
\end{equation}

which does not depend on the parameters of the beam or on the position where the field is evaluated. This result is confirmed in Figure~\ref{fig:Helicity}(a), which shows the normalized helicity density (computed as $h\omega/u$ and evaluated at an arbitrary position with $\rho \leq 20 w_0$ and $z \leq 10 z_R$) versus $\hat{V}$ and $\psi$. The respective maximum and minimum normalized helicity densities are found for $\hat{V}=1$, $\psi = \pm\pi/2$ rad. Figure~\ref{fig:Helicity}(b) denotes the vertical cross-section of Figure~\ref{fig:Helicity}(a) for $\hat{V}=1$, which is identified as the sinusoidal term $\hat{h} = \text{sin}(\psi)$ in Eq.~(\ref{eq:ARPBNormalizedHelicityDensity}). Figure~\ref{fig:Helicity}(c) shows the horizontal cross-section of Figure~\ref{fig:Helicity}(a) for $\psi = \pi/2$, described by $\hat{h} = 2V/(1+V^2)$. The ARPB is optimally chiral ($\hat{h}=\pm 1$) when the APB leads or lags the RPB by a quarter of a period and the beams have amplitudes of the same magnitude, i.e., $V_A = \pm iV_R$, in agreement with what was observed in Ref.~\cite{hanifeh_optimally_2020}. This condition is equivalent to the optimal chirality condition $\mathbf{E}=\pm i \eta_0\mathbf{H}$. Indeed, the fields of an ARPB with $V_A=\pm i V_R$ are 

\begin{equation}
    \begin{array}{c}
         \mathbf{E}_\perp = \frac{\rho }{w^2}f V_R \left[\left(A_\rho + i B_\rho\right)\,\hat{\bm{\rho}} \pm i\,\hat{\bm{\varphi}}\right], \\
         \mathbf{H}_\perp = \mp i \mathbf{E}_\perp / \eta_0 , \\
         E_z = \frac{2i}{kw^2} f V_R\left(A_z + iB_z\right), \\
         H_z = \mp i E_z / \eta_0.
    \end{array}
    \label{eq:OptimallyChiralARPBSimplified}
\end{equation}

Even though the OC-ARPB attains its maximum helicity density on an annular ring around the beam axis (with the same topology as the energy density $u$), the relation $\mathbf{E}= \pm i\eta_0 \mathbf{H}$, and therefore $\hat{h}=\hat{h}_\perp + \hat{h}_z = 1$, hold locally everywhere in space. 

\begin{figure*}[t]%
\centering
\subfloat[]{\includegraphics[width=0.33\linewidth]{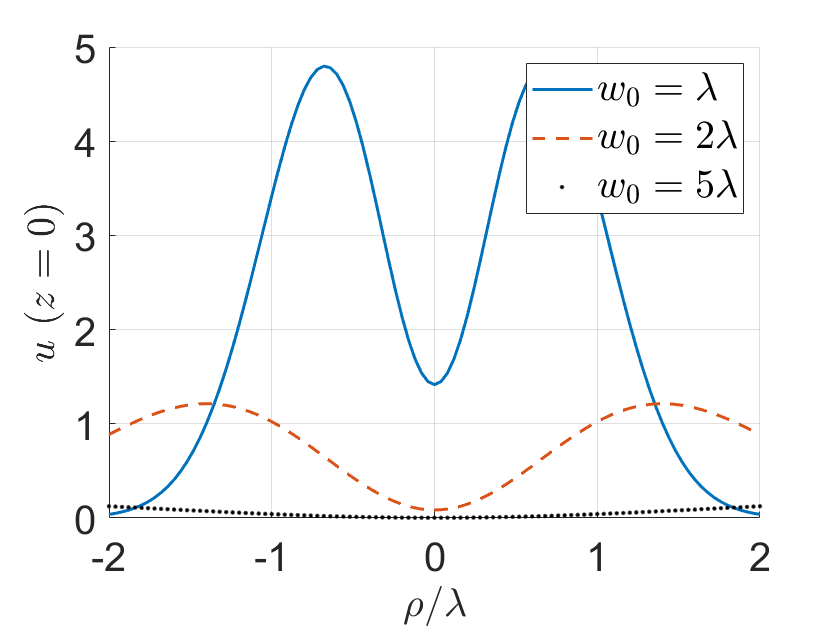}}
\subfloat[]{\includegraphics[width=0.33\linewidth]{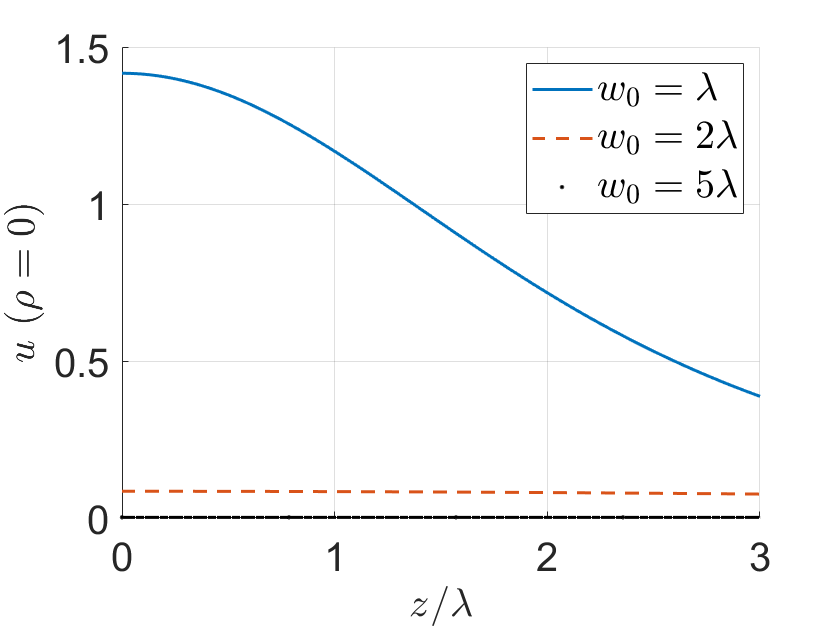}}
\subfloat[]{\includegraphics[width=0.33\linewidth]{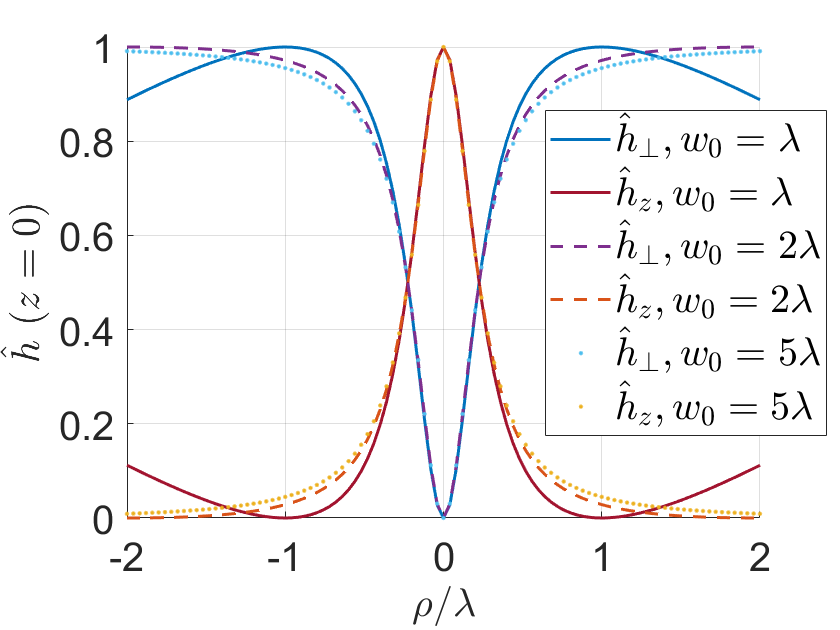}}
\caption[\linewidth]{Comparison of the energy $u$ and helicity $h$ densities of an OC-ARPB (of $1$ mW power) with $w_0=\lambda$, $w_0=2\lambda$, and $w_0=5\lambda$. (a) Depicts the energy density at focus in the transverse plane and (b) shows the energy density on the beam axis. In both figures, the energy density increases as the ratio $w_0/\lambda$ decreases. (c) Depicts the transverse $\hat{h}_\perp$ and longitudinal $\hat{z}$ normalized helicity density of the three OC-ARPB. Even though the focused ARPB has a higher helicity density, the region of space where its longitudinal component dominates over the transverse one remains approximately constant at $\rho\approx 0.225 \lambda$. }
\label{fig:HelicityComponents}
\end{figure*}

As discussed in Section~\ref{ch:ARPB}, when $w_0$ decreases, the relative increase in the magnitude of the longitudinal fields compared to the transverse fields becomes more pronounced. Therefore, focusing an ARPB through a lens increases its energy density, and consequently, the upper bound of its chirality. Figure~\ref{fig:HelicityComponents} illustrates this behavior by comparing the energy densities and the normalized helicity densities of two OC-ARPBs of $1$ mW power with $w_0=\lambda$ and $w_0=2\lambda$, where $\lambda=400$ nm. Figures~\ref{fig:HelicityComponents}(a) and (b) show the energy density $u$ of both cases, at the focus and on the $z$ axis, respectively. Focusing the ARPB by decreasing the ratio $w_0/\lambda$ increases the energy density and, consequently, the helicity density. Figure~\ref{fig:HelicityComponents}(c) displays the normalized helicity densities of the transverse $\hat{h}_\perp$ and longitudinal $\hat{h}_z$ components for both beams. Although the focused OC-ARPB has a larger helicity density, the region where the longitudinal component dominates over the transverse component remains roughly the same, at approximately $\rho\approx 0.225\lambda$.

As a result of the collinearity between the electric and magnetic field phasors, OCL displays an exceptional electric-magnetic symmetry in its energy and spin densities, i.e., $u_e = u_m$ and $\bm{\sigma}_e = \bm{\sigma}_m$ \cite{hanifeh_optimally_2020}. The Poynting vector $\mathbf{S}^{\text{OC-ARPB}}$ of an OC-ARPB is

\begin{equation}
    \begin{array}{c}
    \mathbf{S}^{\text{OC-ARPB}} = S_0 \left[\right.2B_z\hat{\bm{\rho}}
    \mp 2\left(A_\rho A_z + B_\rho B_z\right)\hat{\bm{\varphi}} + k\rho A_\rho\hat{\bm{z}}\left.\right],
    \end{array}
    \label{eq:MomentumOptimallyChiralARPB}
\end{equation}

which is real-valued globally and, consequently, all the energy flux density contributes to the linear momentum density; an OC-ARPB has no reactive energy flux density. Moreover, the Poynting vector of an OC beam is collinear with its SAM density, as demonstrated in the Supplementary Information of Ref.~\cite{hanifeh_optimally_2020}. For an OC-ARPB, the spin density $\bm{\sigma}^{\text{OC-ARPB}}$ is

\begin{equation}
    \begin{array}{c}
        \bm{\sigma}^{\mathrm{ARPB}}= \sigma_0\left[\right.\pm2B_z\hat{\bm{\rho}} -2\left(A_\rho A_z + B_\rho B_z\right)\hat{\bm{\varphi}} \pm k\rho A_\rho\hat{\bm{z}}\left.\right].
    \end{array}
    \label{eq:SpinDensitiesOptimal}
\end{equation}

\begin{figure}[t]%
\centering
\subfloat[]{\includegraphics[width=0.36\textwidth]{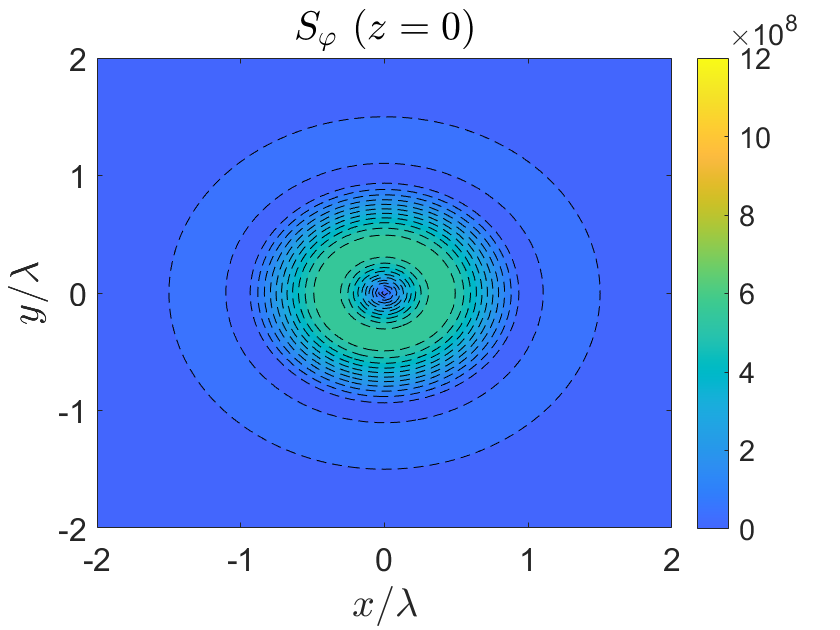}}
\hfill
\subfloat[]{\includegraphics[width=0.36\textwidth]{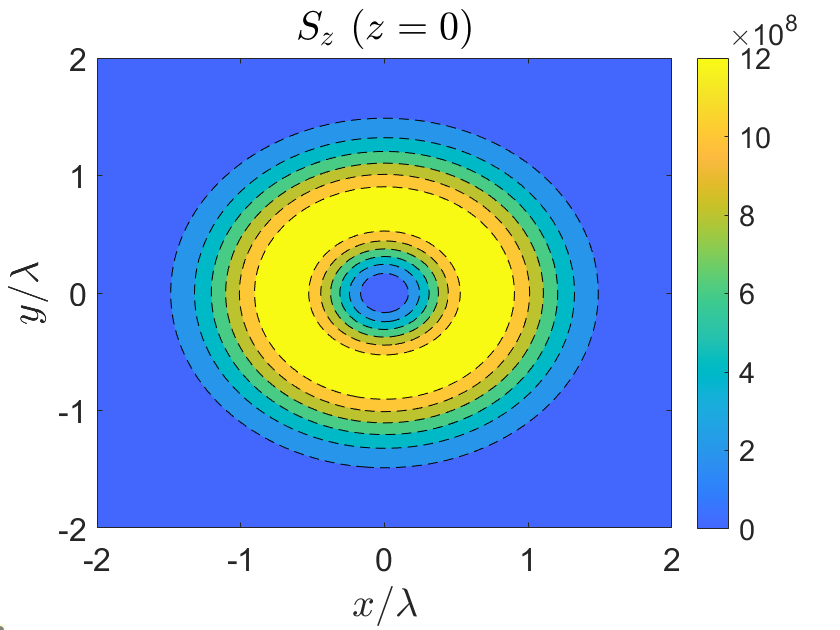}}
\caption[\linewidth]{Magnitudes of the (a) azimuthal and (b) longitudinal components of the Poynting vector $\mathbf{S}$ of an OC-ARPB (of $1$ mW power at $\lambda = 400$ nm) in the transverse plane at the focus of the beam, i.e., $z=0$. The radial component $S_\rho$ has not been plotted since it is practically zero at the focus of the beam. The azimuthal component $S_\varphi$ peaks in a small ring close to the beam axis, since it is associated with the radial and longitudinal fields. The longitudinal component $S_z$ is the largest of them since it is only associated with the transverse fields, which are significantly larger than the longitudinal components. Due to azimuthal symmetry and vanishing transverse fields on the beam axis, the components of the Poynting vector describe an annular ring around it.}
\label{fig:OC-Sigma}
\end{figure}

As demonstrated here for the OC-ARPB, although it is easily verified that it applies for all OCL, $\mathbf{S}= \pm \omega c \bm{\sigma}$. Since the spin density is a real quantity, OCL has a purely real energy flux density $\mathbf{S}\in\Re$, and therefore a maximized linear momentum density $\mathbf{p}=\pm k\bm{\sigma}$, which changes sign for $\psi=+ \pi/2$ or $\psi=- \pi/2$. The normalization constants of the Poynting vector and the spin density of the ARPB also follow the same relation $S_0 = \omega c \sigma_0$. Figure~\ref{fig:OC-Sigma} displays the magnitudes of the (a) azimuthal and (b) longitudinal components of the Poynting vector of an OC-ARPB $\mathbf{S}^{\text{OC-ARPB}}$ on the transverse plane at the beam focus ($z=0$). Its radial component $S_\rho$, which is approximately zero near the beam's focus (see Eq.~(\ref{eq:MomentumOptimallyChiralARPB}), where $B_z \approx 0$). The azimuthal component $S_\varphi$ has a peak described by a small ring around the beam axis. This occurs because it is partly associated with the longitudinal fields of the OC-ARPB, which peak on the beam axis. The longitudinal component of the Poynting vector $S_z$ is associated with the transverse fields and therefore describes a bigger ring around the beam axis.

In summary, the OC-ARPB displays the enhanced chirality-probing capabilities characteristic of OCL, together with an exceptional spatial separation of longitudinal and transverse properties. Unlike CPL, which only measures transverse chirality, the ARPB measures exclusively longitudinal chirality along the beam axis, making it preferable because transverse chirality measurements can be mistaken for sample anisotropy \cite{kamandi_unscrambling_2018}. Using the ARPB beam as well as some other beams with longitudinal axial chirality, such as those discussed in Refs.~\cite{koksal_optical_2022, koksal_hopf_2022, babiker_zero_2022, wozniak_interaction_2019, forbes_optical_2022, forbes_enantioselective_2022} whose chirality is proportional to the topological charge $l$, eliminate the chiral-anisotropy ambiguity issue by only measuring the chirality of a particle located on the beam axis. The advantage of the OC-ARPB over these beams is that its chirality can be easily tuned by adjusting the phase parameter $\psi$, as shown in Section~\ref{ch:GenerationManipulation}, and therefore it is a flexible and versatile option for achieving controlled probing or enantioseparation of chiral nanoparticles.

\begin{figure*}[t]%
\centering
\subfloat[]{\includegraphics[width=0.4\textwidth]{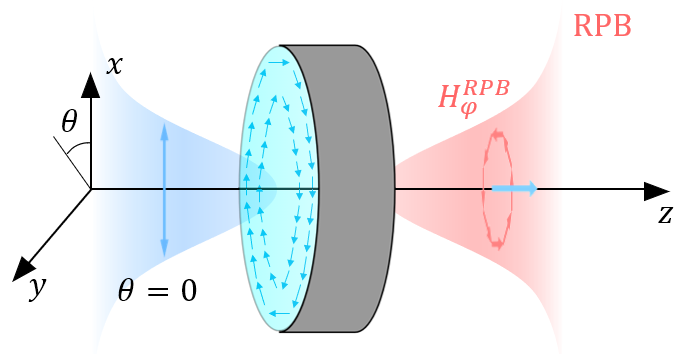}}
\subfloat[]{\includegraphics[width=0.4\textwidth]{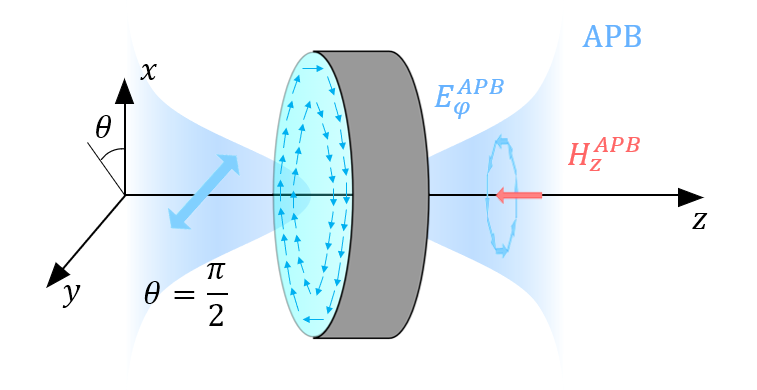}}
\hfill
\subfloat[]{\includegraphics[width=0.4\textwidth]{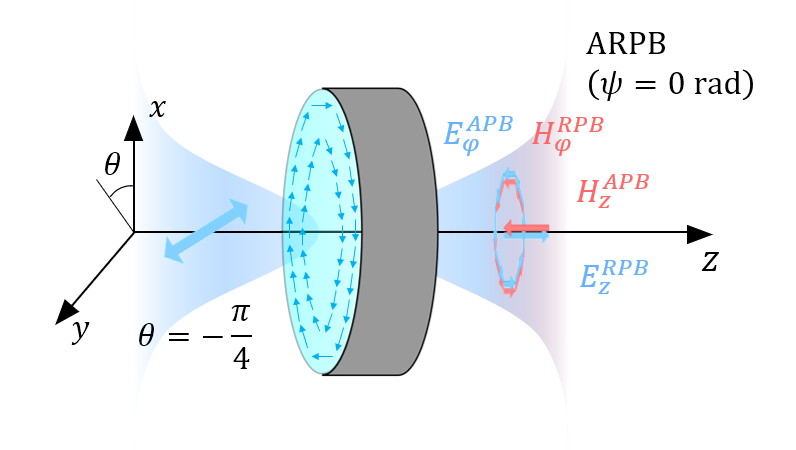}}
\subfloat[]{\includegraphics[width=0.4\textwidth]{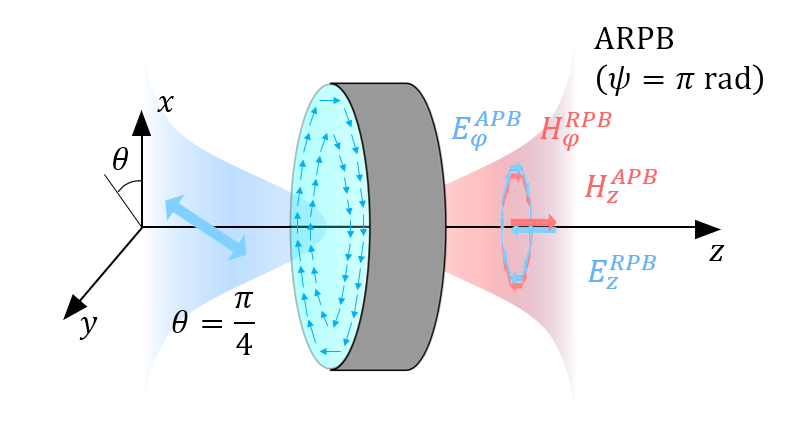}}
\caption{Schematic of a linearly polarized beam traveling through an SWP. (a) An $x$-polarized beam is transformed into an RPB. (b) An $y$-polarized beam is transformed into an APB. (c) A beam polarized at $\theta = \pi/4$, with $\theta$ the angle between the $x$ and $y$ axis, is transformed into the sum of an RPB and an APB with opposite phases. (d) A beam polarized at $\theta = -\pi / 4$ is transformed into a sum of an RPB and an APB with the same phase.}
\label{fig:SWP}
\end{figure*}

\section{ARPB generation}
\label{ch:GenerationManipulation}

We show an optical setup that generates an ARPB with arbitrary $\hat{V}$ and $\psi$ from an arbitrarily polarized laser source. An option to transform familiar wave polarization types (such as linear, circular, and elliptical) into vector modes (such as radial or azimuthal) is the use of an S-waveplate (SWP). The SWP is a combination of half-wave plate (HWP) segments whose slow axis is changed to maintain an azimuthal angle with respect to a reference axis \cite{bhargava_polarization_2017}, and can produce an APB, an RPB, or a combination of both by selective input polarization \cite{lin_cladding_2014}. Its Jones matrix is \cite{matijosius_formation_2014}

\begin{equation}
        \underline{\mathbf{H}}_{S}=\left(\begin{array}{cc}
        \cos\varphi & \sin\varphi \\
        \sin\varphi & -\cos\varphi
    \end{array}\right).
    \label{eq:JonesSwaveplate}
\end{equation}

\begin{table}[H]
    \centering
    \begin{tabular}{|ccc|}
    \hline
     \(\mathbf{J}_{in}\) & $\to$ & \(\mathbf{J}_{out}= \underline{\mathbf{H}}_{S}\mathbf{J}_{in} \) \\
     \hline
     \( \left( 1, \ 0 \right)^T\) & $\to$ & \(\mathbf{J}_{RPB} = \left(\cos(\varphi), \  \sin(\varphi)\right)^T \) \\
     \hline
     \( \left( 0, \ -1 \right)^T\) & $\to$ & \(\mathbf{J}_{APB} = \left(-\sin(\varphi), \  \cos(\varphi)\right)^T \) \\
     \hline
     \( \frac{1}{\sqrt{2}}\left(1, \  -e^{i\psi}\right)^T\) & $\to$ & \(\mathbf{J}_{ARPB} = \mathbf{J}_{RPB} + e^{i\psi}\mathbf{J}_{APB} \),\\
     \hline
    \end{tabular}
    \caption{SWP relevant input and output pairs. It provides the inputs required to obtain an RPB, an APB, and an ARPB with $\hat{V}=1$ and arbitrary $\psi$, respectively. The superscript $T$ denotes the transpose operator and $\varphi$ is the azimuthal coordinate.}
    \label{tab:InputOutputSWP}
\end{table}

\begin{figure}[h]
    \centering
    \includegraphics[width = 0.4\textwidth]{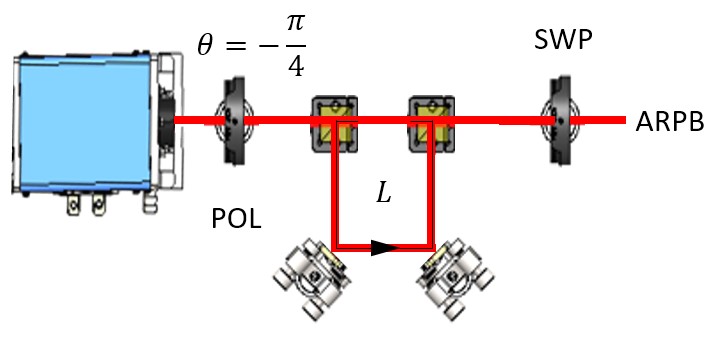}
    \caption{Optical setup to obtain an ARPB with $\hat{V}=1$ and arbitrary $\psi$. The beam is polarized at an angle $\theta = - \pi / 4$ with respect to the $x$-axis, and the $y$-component is added a phase-shift $\psi$ using a delay arm of length $L$. The resulting beam travels through an SWP, resulting in the desired ARPB.}
    \label{fig:Setup}
\end{figure}

The Jones vectors for relevant input/output pairs are summarized in Table~\ref{tab:InputOutputSWP} and depicted in Figure~\ref{fig:SWP}. In particular, Figure~\ref{fig:SWP}(a) shows that an $x$-polarized beam becomes an RPB, and Figure~\ref{fig:SWP}(b) that a $y$-polarized beam becomes an APB. Figure~\ref{fig:SWP}(c) and (d) illustrate the conversion of linearly polarized beams at angles $\theta = \mp \pi/4$ with respect to the $x$ axis into achiral ARPBs with $\hat{V}=1$ and either $\psi = 0$ or $\psi = \pi$, respectively. An ARPB with arbitrary $\hat{V}$ and $\psi$ can be obtained by setting a linear polarizer at an angle $\theta$ from the $x$ axis and adding a phase delay equal to $\psi$ on the $y$-polarized component through a delay arm of length $L$, as shown in Figure~\ref{fig:Setup}. In this arrangement, the amplitude ratio $\hat{V}$ between the APB and the RPB is controlled by the angle $\theta$ of the linear polarizer relative to the $x$ axis, which is given by

\begin{equation}
	\hat{V}=|\tan(\theta)|.
	\label{eq:vtheta}
\end{equation}

To obtain $\hat{V}=1$, the linear polarizer must be set to $\theta = \pm\pi/4$, where choosing the minus sign removes a $180\degree$ phase from the resulting APB upon conversion by the SWP. The length $L$ of the delay arm sets the phase parameter $\psi$, and the two are related by the equation

\begin{equation}
    kL - \tan^{-1}\left(\frac{L}{z_R} \right) = \psi,
    \label{eq:DelayArmLength}
\end{equation}

which depends on $\lambda$ and $w_0$. 

An alternative method for introducing the phase delay $\psi$ between the $y$- and $x$-polarized components of a linearly polarized beam relies on using a HWP and a quarter-wave plate (QWP). The Jones matrix of a HWP at angle $\theta_h$ with respect to the $x$ axis is \cite{theocaris_matrix_1979} 

\begin{equation}
    \underline{\mathbf{HWP}}(\theta_h)= \left(\begin{array}{cc}
        -\cos(2\theta_h) & -\sin(2\theta_h) \\
        -\sin(2\theta_h) & \cos(2\theta_h)
    \end{array}\right), 
    \label{eq:HWP}
\end{equation}

and that of a QWP at angle $\theta_Q$ is 

\begin{equation}
    \underline{\mathbf{QWP}}(\theta_q)= \left(\begin{array}{cc}
        i\cos^2\theta_q + \sin^2\theta_q & (i-1)\cos\theta_q\sin\theta_q \\
        (i-1)\cos\theta_q\sin\theta_q & i\sin^2\theta_q + \cos^2\theta_q
        \end{array}\right).
    \label{eq:QWP}
\end{equation}

The resulting transformation reads

\begin{equation}
    \mathbf{J}_{out} = \underline{\mathbf{HWP}}(\theta_h) \underline{\mathbf{QWP}}(\theta_q) \mathbf{J}_{in}.
    \label{eq:PhaseShifter}
\end{equation}

\begin{figure}[h]
    \centering
    \includegraphics[width = 0.4\textwidth]{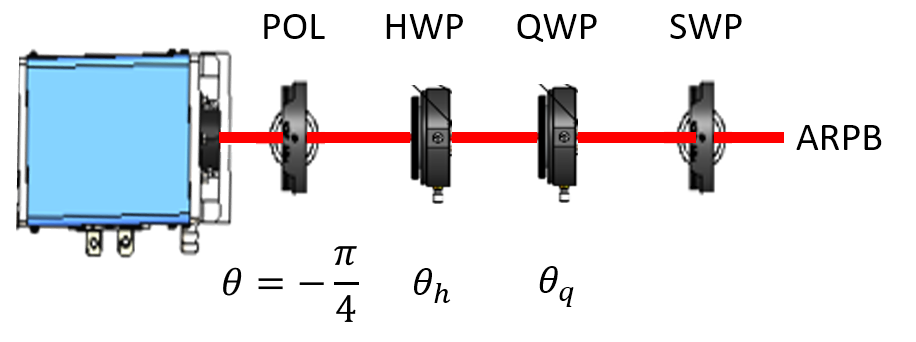}
    \caption{Optical setup to obtain an ARPB with $\hat{V}=1$ and arbitrary $\psi$ using a QWP and a HWP at angles $\theta_q$ and $\theta_h$, respectively, instead of the delay arm from Figure~\ref{fig:Setup}. Possible angle combinations that provide the most relevant $\psi$ values are shown in Table \ref{tab:AnglesQWPHWPShift}.}
    \label{fig:PhaseShifter}
\end{figure}

The angle combinations that provide the $\psi$ phase delay depend on the initial and final position of the polarization on the Poincaré sphere (see Ref.~\cite{theocaris_matrix_1979} for details). Replacing the delay arm in Figure~\ref{fig:Setup} with the HWP and the QHP, as depicted in Figure~\ref{fig:PhaseShifter}, indicates that the incident polarization is $\mathbf{J}_{in} = (1, \ 1)^T$ and the final is $\mathbf{J}_{out} = (1, \ e^{i\psi})^T$. Examples of angle combinations for the most relevant $\psi$ values for this transformation are shown in Table~\ref{tab:AnglesQWPHWPShift}.

\begin{table}[H]
    \centering
    \begin{tabular}{|ccc|}
    \hline
     $\psi$ (rad)  & $\theta_H$ (rad) & $\theta_Q$ (rad) \\
     \hline
     $0$ &  $7\pi/4$ & $5\pi/4$ \\
     \hline
     $\pi/2$ & $0$ & $0$ \\
     \hline
     $\pi$ & $3\pi/2$ & $5\pi/4$ \\
     \hline
     $-\pi/2$ & $0$  & $-\pi/2$ \\
     \hline
    \end{tabular}
    \caption{Examples of angle combinations that provide the most relevant phase delay $\psi$ values for the polarization transformation of a beam with $\mathbf{J}_{in} = (1, \ 1)^T$ into a beam with $\mathbf{J}_{out} = (1, \ e^{i\psi})^T$.}
    \label{tab:AnglesQWPHWPShift}
\end{table}

Using two cascaded HWPs at angles $\theta_1 = \delta/2$ and $\theta_2 = 0$ rotates the polarization by an angle $\delta$ with respect to the input vector \cite{mcdonald_matrix_2016}. The Jones matrix of the resulting polarization rotator, i.e., $\underline{\mathbf{PR}}(\delta)=\underline{\mathbf{HWP}}(\delta/2)\underline{\mathbf{HWP}}(0)$, is

\begin{equation}
    \underline{\mathbf{PR}}(\delta) = 
    \left(\begin{array}{cc}
        \cos\delta & -\sin\delta \\
        \sin\delta & \cos\delta
    \end{array}\right).
    \label{eq:PolRotMatrix}
\end{equation}

\begin{figure}[H]
    \centering
    \includegraphics[width = 0.4\textwidth]{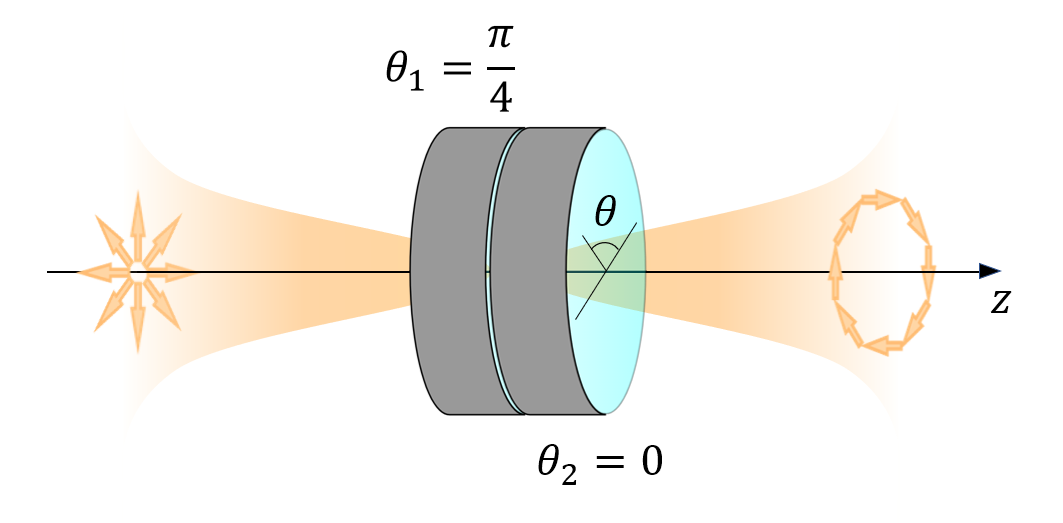}
    \caption{Schematic for a radial-to-azimuthal polarization rotator using two HWPs with an angle difference of $\theta_1 - \theta_2 = \pi/4$.}
    \label{fig:PolarizationRotator}
\end{figure}

This device can be adapted for a variety of purposes. For example, a radial-to-azimuthal polarization rotator is obtained by setting $\delta = \pi/2$, as depicted in Fig.~\ref{fig:PolarizationRotator}. The order of the HWPs is inverted in Ref.~\cite{moreno_polarization_2012}, which adds a $\pi$ phase to the resulting APB. Whilst this phase difference is not relevant in measurements, it may affect light manipulation. Using the polarization rotator with $\delta = \pi$ instead changes the handedness of an incident OC-ARPB.

Although the main focus in this section is on generating both configurations of the OC-ARPB for chirality probing and enantioseparation purposes, the setups depicted in Figure~\ref{fig:Setup} and \ref{fig:PhaseShifter} can also be used to smoothly adjust the OAM density of the ARPB, either by varying the length $L$ of the delay arm or the angles of the wave plates, which has practical applications in conducting OAM-related experiments.

\section{Conclusion}
\label{ch:Conclusion}

We have introduced a concise notation to describe the APB, the RPB, and the ARPB. We have investigated the properties of the ARPB and derived the analytical expressions for its field quantities. Our findings demonstrate that an in-phase superposition of an APB and an RPB results in an achiral ARPB. An out-of-phase superposition, however, results in a chiral ARPB with larger linear momentum and spin densities (both in terms of magnitude and number of components) than those of achiral beams. Moreover, a chiral ARPB carries a longitudinal OAM density despite having no azimuthal phase dependence, which can be continuously adjusted by varying the phase delay $\psi$ between the APB and the RPB. Indeed, the ARPB can be {\em optimally chiral} and exhibits maximum linear momentum, angular momentum, and helicity densities. The transverse fields of the ARPB vanish on the beam axis, and therefore the same occurs to its linear and angular momentum densities, where only the energy and helicity densities associated with the longitudinal fields persist. We have shown that focusing an ARPB through a lens boosts its longitudinal fields more than the transverse ones, which results in an enhanced longitudinal chirality. Therefore, the ARPB provides a controlled environment that may be used to trap and probe chiral nanoparticles. Accordingly, the next step would be to calculate the photoinduced forces on a chiral dipolar particle illuminated by an ARPB, with an emphasis on the optimally chiral configuration. To aid future research experiments, we have also provided the schematics for an optical setup that generates an ARPB and provided a polarization rotator that changes the handedness of an OC-ARPB.

\ifCLASSOPTIONcaptionsoff
  \newpage
\fi

\bibliographystyle{IEEEtran}
\bibliography{IEEEabrv,references}

\end{document}


\maketitle
\renewcommand{\theequation}{S.\arabic{equation}}
\renewcommand{\thefigure}{S.\arabic{figure}}

\section*{A: APB electric field in different polarization basis}

Expressing the electric field $\mathbf{E^{\mathrm{APB}}}$ of the azimuthally polarized beam (APB) in different polarization bases can provide valuable information about the field's topology, especially with regards to its orbital angular momentum (OAM) density. As per Eq.~(1) in the manuscript, the APB electric field is purely azimuthal, i.e., $E_\varphi^{\mathrm{APB}}$, which ultimately results in the presence of a non-zero azimuthal magnetic SAM density $\sigma_{m,\varphi}$, as confirmed in Section B of the Supporting Information. When expressed in the circular polarization basis, $\mathbf{E^{\mathrm{APB}}}$ is given by

\begin{equation}
    \mathbf{E^{\mathrm{APB}}}=\frac{i\sqrt{2}}{2}\frac{V_A\rho}{w^2}f\left(e^{i\varphi}\,\hat{\bm{e}}_{\text{RH}}-e^{-i\varphi}\,\hat{\bm{e}}_{\text{LH}}\right),
\label{eq:APBCPLBasis}
\end{equation}

where the circular unitary vectors are $\hat{\bm{e}}_{\text{RH}} = \left(\hat{\bm{x}}-i\hat{\bm{y}}\right)/\sqrt{2}$ for right-circularly polarized light
and $\hat{\bm{e}}_{\text{LH}} = \left(\hat{\bm{x}}+i\hat{\bm{y}}\right)/\sqrt{2}$ for left-circularly polarized light \cite{hanifeh_optimally_2020}. The APB is described in Eq.~(\ref{eq:APBCPLBasis}) as a composite of two phase-shifted circularly polarized vortex beams of opposite handedness and OAM density. Writing the electric field in terms of the Laguerre-Gaussian (LG) modes $u_{\pm1,0}$, as defined in Eq.~(4) in the manuscript, results in 

\begin{equation}
	\mathbf{E^{\mathrm{APB}}}=\frac{i\sqrt{2}}{2}V_{\mathrm{A}}\left(u_	{1,0}\,\hat{\bm{e}}_{\text{RH}}-u_{-1,0}\,\hat{\bm{e}}_{\text{LH}}\right)e^{ikz},
	\label{eq:APBCPLBasisLG}
\end{equation}

as shown in Eq.~(13.2) in Ref.~\cite{jiang_theory_2021}. Even though the APB electric field is comprised of two vortex beams, which display their characteristic azimuthal phase dependence $e^{il\varphi}$ \cite{leach_vortex_2005}, it does not carry any OAM density. The local cancellation of the OAM density is clear when $\mathbf{E^{\mathrm{APB}}}$ is written in the cartesian coordinate system, i.e., 

\begin{equation}
    \mathbf{E^{\mathrm{APB}}}=\frac{V_A\rho}{w^2}f\left(-\sin{\varphi}\,\hat{\bm{x}}+\cos{\varphi}\,\hat{\bm{y}}\right).
    \label{eq:APBSinCos}
\end{equation}

Due to the duality between the APB and the RPB, Eq.~(\ref{eq:APBCPLBasis}) and~(\ref{eq:APBSinCos}) apply for $\mathbf{H}^{\mathrm{RPB}}$ with $V_R / \eta_0$ instead of $V_A$. Thus, the RPB does not have a phase variation around the axis, despite consisting of a phase-shifted combination of two CP beams with $e^{\pm i\varphi}$.

\section*{B: Cycle-averaged field quantities for the APB, the RPB, and the ARPB}

The cycle-averaged energy density of an electromagnetic (EM) field is $u = u_e + u_m = \varepsilon_0 |\mathbf{E}|^2/ 4 + \mu_0|\mathbf{H}|^2/4$, in units of $\left(\text{J} / \text{m}^3 \right)$ \cite{angelsky_structured_2020}. The energy densities of the APB and the RPB are

\begin{equation}
        \begin{array}{c}
        u_e^{\mathrm{APB}} = \frac{1}{2}u_0\hat{V}^2(k\rho)^2, \\
        u_m^{\mathrm{APB}} = \frac{1}{2}u_0\hat{V}^2\left[(k\rho)^2\left(A_\rho^2 + B_\rho^2\right) +4\left(A_z^2 + B_z^2\right)\right], \\
        u_e^{\mathrm{RPB}} = \frac{1}{2}u_0\left[(k\rho)^2\left(A_\rho^2 + B_\rho^2\right) +4\left(A_z^2 + B_z^2\right)\right], \\
        u_m^{\mathrm{RPB}} = \frac{1}{2}u_0(k\rho)^2,
        \end{array}
    \label{eq:Energy}
\end{equation}

where $u_0 = \frac{\varepsilon_0|f|^2}{2k^2w^4}|V_R|^2$ as defined in the manuscript. The ARPB energy densities $u_e^{\mathrm{ARPB}}$ and $u_m^{\mathrm{ARPB}}$ are the sum of the energy densities of the APB and the RPB. The complex Poynting vector is $\mathbf{S}=\left(\mathbf{E}\times\mathbf{H}^*\right) / 2$ as in Ref.~\cite{hanifeh_optimally_2020}, with units of $\left(\text{W} / \text{m}^{2}\right)$. Its real part represents the time-averaged energy flux density \cite{hanifeh_optimally_2020} and is related to the linear momentum of the field, i.e., $\mathbf{p}=\Re(\mathbf{S})/c^2$ \cite{bliokh_extraordinary_2014}. Its imaginary component is the reactive energy flux density of the beam, which represents the oscillating flow of the energy density through a surface. The Poynting vectors of the three beams under consideration are

\begin{equation}
    \begin{array}{c}
        \mathbf{S}^{\mathrm{APB}} = \frac{1}{2}S_0\hat{V}^2\left[2i\left(A_z-iB_z\right)\,\hat{\bm{\rho}} + k\rho \left(A_\rho -iB_\rho\right)\,\hat{\bm{z}} \right], \\
        \mathbf{S}^{\mathrm{RPB}} = \frac{1}{2}S_0\left[-2i\left(A_z+iB_z\right)\,\hat{\bm{\rho}} + k\rho\left(A_\rho +iB_\rho\right)\,\hat{\bm{z}} \right], \\
        \mathbf{S}^{\mathrm{ARPB}} = \mathbf{S}^{\mathrm{APB}} + \mathbf{S}^{\mathrm{RPB}} - 2S_0\hat{V}\left(\sin\psi+i\cos\psi\right)\left(A_\rho A_z + B_\rho B_z\right)\,\hat{\bm{\varphi}},
    \end{array}
    \label{eq:Momentum}
\end{equation}

where $S_0 = \frac{\rho|f|^2}{k\eta_0w^4}|V_R|^2= 2\rho \omega u_0$. The radial components of the Poynting vector $S_\rho$ of the APB and RPB are mostly imaginary near the axis and focus of the beam (where $B_z \approx 0$ and $A_z \approx 1$) and have a $\pi$ phase shift. Consequently, the flux of energy density of both beams is mostly reactive, and it oscillates in opposite directions for the APB and the RPB. The absence of an azimuthal component Poynting vector of the individual APB and the RPB reinforces the notion that these beams do not carry any longitudinal OAM density. In contrast, the azimuthal component of the ARPB Poynting vector $S_\varphi^{\mathrm{ARPB}}$ is different than zero. This component is purely imaginary for an achiral ARPB (with $\psi=0,\pi$), and as a result, it does not contribute to the linear momentum density of the beam. On the other hand, in the case of a chiral ARPB (with $\psi \neq 0, \pi$), the ratio of active to reactive energy flux density in an ARPB is determined by the tangent of the phase parameter $\psi$. In the case of optimally chiral light (with $\psi = \pm\pi/2$), the ARPB has a purely real $S_\varphi^{\mathrm{ARPB}}$, and all the energy flux density contributes to the azimuthal linear momentum density, which changes sign for $\psi=+ \pi/2$ or $\psi=- \pi/2$. The total powers of the APB, the RPB, and the ARPB in the $z$ direction are 

\begin{equation}
    \begin{array}{c}
         P^{\mathrm{APB}}=\frac{|V_A|^2}{2\eta_0}\left(1-\frac{1}{\omega z_R}\right),\\
         P^{\mathrm{RPB}}=\frac{|V_R|^2}{2\eta_0}\left(1-\frac{1}{\omega z_R}\right), \\
         P^{\mathrm{ARPB}} = P^{\mathrm{APB}} + P^{\mathrm{RPB}},
    \end{array}
    \label{eq:Power}
\end{equation}

where $P^{\mathrm{APB}}$ is calculated in Ref.~\cite{veysi_focused_2016}, $P^{\mathrm{RPB}}$ is obtained from the duality between these beams, described in Eq.~(5) in the manuscript, and the power of the ARPB is the sum of the two underlying beams. 

The cycle-averaged spin angular momentum (SAM) density is given by $\bm{\sigma} = \bm{\sigma}_e + \bm{\sigma}_m = -\frac{\varepsilon_0}{4i\omega }\left(\mathbf{E}\times\mathbf{E}^*\right)-\frac{\mu_0}{4i\omega }\left(\mathbf{H}\times\mathbf{H}^*\right)$ \cite{hanifeh_optimally_2020}, and it has units of angular momentum over volume, i.e., $\left(\text{Nms} / \text{m}^3  = \text{Ns} / \text{m}^2 \right)$. The SAM densities associated with the three beams under study are

\begin{equation}
        \begin{array}{c}
            \bm{\sigma}_e^{\mathrm{APB}} = 0, \\
            \bm{\sigma}_m^{\mathrm{APB}} = -\sigma_0\hat{V}^2\left(A_\rho A_z + B_\rho B_z\right)\,\hat{\bm{\varphi}}, \\
            \bm{\sigma}_e^{\mathrm{RPB}} = -\sigma_0\left(A_\rho A_z + B_\rho B_z\right)\,\hat{\bm{\varphi}}, \\
            \bm{\sigma}_m^{\mathrm{RPB}} = 0, \\
            \bm{\sigma}_e^{\mathrm{ARPB}} =\frac{1}{2}\sigma_0\hat{V}\left[\right.2\left(A_z\cos\psi+B_z\sin\psi\right)\,\hat{\bm{\rho}}- k\rho \left(B_\rho\cos\psi-A_\rho\sin\psi\right)\,\hat{\bm{z}}\left.\right] + \bm{\sigma}_e^{\mathrm{RPB}}, \\
            \bm{\sigma}_m^{\mathrm{ARPB}}=-\frac{1}{2}\sigma_0\hat{V}\left[\right.2\left(A_z\cos\psi-B_z\sin\psi\right)\,\hat{\bm{\rho}}- k\rho \left(B_\rho\cos\psi+A_\rho\sin\psi\right)\,\hat{\bm{z}}\left.\right] + \bm{\sigma}_m^{\mathrm{APB}},
        \end{array}
        \label{eq:Spin}
\end{equation}

where $\sigma_0 = \frac{\varepsilon_0\rho|f|^2}{\omega k w^4}|V_R|^2=S_0/(\omega c)$. The APB and RPB have purely transverse spin densities (TSD), which arise in laterally-confined light \cite{neugebauer_magnetic_2018} and have applications in chiral light-matter interactions \cite{wang_lateral_2014, hayat_lateral_2015}. The ARPB, however, also has radial and longitudinal spin densities. They arise from the interaction between the APB and the RPB and hence depend on the phase parameter $\psi$. 

In Ref.~\cite{bliokh_extraordinary_2014}, the authors introduce the total angular momentum density $\mathbf{j}=\mathbf{r}\times\mathbf{p}=\mathbf{l} + \bm{\sigma}$. Therefore, the OAM density $\mathbf{l}$ can be calculated from the linear momentum $\mathbf{p}$ and the SAM $\bm{\sigma}$ densities as

\begin{equation}
    \mathbf{l}=\mathbf{r}\times\mathbf{p} - \bm{\sigma}.
    \label{eq:OAMfromLinearSAMGeneric}
\end{equation}

In cylindrical coordinates, we obtain

\begin{equation}
    \begin{array}{c}
         l_\rho = -p_\varphi z - \sigma_\rho, \\
         l_\varphi = p_\rho z - p_z \rho - \sigma_\varphi, \\
         l_z = p_\varphi \rho - \sigma_z.
    \end{array}
    \label{eq:OAMfromLinearSAM}
\end{equation}

which applies to any of the APB, RRB, and ARPB cases. While the relation between the azimuthal component of the linear momentum density and the longitudinal orbital angular momentum (OAM) density $l_z$ is well established \cite{spereits_waves_2013, barnett_natures_2016}, here we have introduced its calculation in terms of the linear momentum and spin densities, whose calculation is straightforward and preferable to calculations involving the gradient of the field vectors. The APB and the RPB have no azimuthal linear momentum density and therefore carry no longitudinal OAM density. However, in the case of a chiral APRB where the APB and the RPB are phase-shifted ($\psi\neq 0, \pi$), there is an energy rotation around the beam axis (due to the asymmetry of Maxwell's equations), which results in a non-zero longitudinal OAM density $l_z^{\mathrm{ARPB}}$ as shown in Eq.~(9) in the manuscript.

The cycle-averaged helicity density $h$ of EM fields is a measure of their chirality. For a freely-propagating monochromatic beam, $h =\Im\left(\mathbf{E}\cdot\mathbf{H}^*\right) /(2\omega c)$ \cite{hanifeh_optimally_2020}. The APB and the RPB are achiral, while the helicity density of the ARPB is

\begin{equation}
    \begin{array}{c}
        h^{\mathrm{ARPB}}=h_0\hat{V}\sin(\psi)\left[(k\rho)^2\left(1 + A_\rho^2 + B_\rho^2\right)+4\left(A_z^2 + B_z^2\right)\right], 
    \end{array}
    \label{eq:ARPBelectricHelicity}
\end{equation}

where $h_0 = \frac{\varepsilon_0|f|^2}{2\omega k^2w^4}|V_R|^2 $. The phase parameter $\psi$ mediates the interaction between the fields of the APB and those of the RPB. For $\psi = 0, \pi$, the interaction between the fields of both beams is minimum, and the energy, linear momentum, OAM, SAM, and helicity densities of the resulting (achiral) ARPB are the sum of those of the APB and the RPB. For $\psi \neq 0,\pi$, these field quantities increase both in magnitude and in the number of non-zero components, which reach a maximum for OCL ($\psi = \pm\pi/2$). Ultimately, the phase parameter $\psi$ is the source of both the longitudinal orbital angular momentum density $l_z^{\mathrm{ARPB}}$ and the helicity density $h^{\mathrm{ARPB}}$, which are proportional in the case of the ARPB. 

\section*{C: APB and RPB field distribution for collimated beams}

\begin{figure}[H]%
\centering
\subfloat[]{\includegraphics[width=0.34\linewidth]{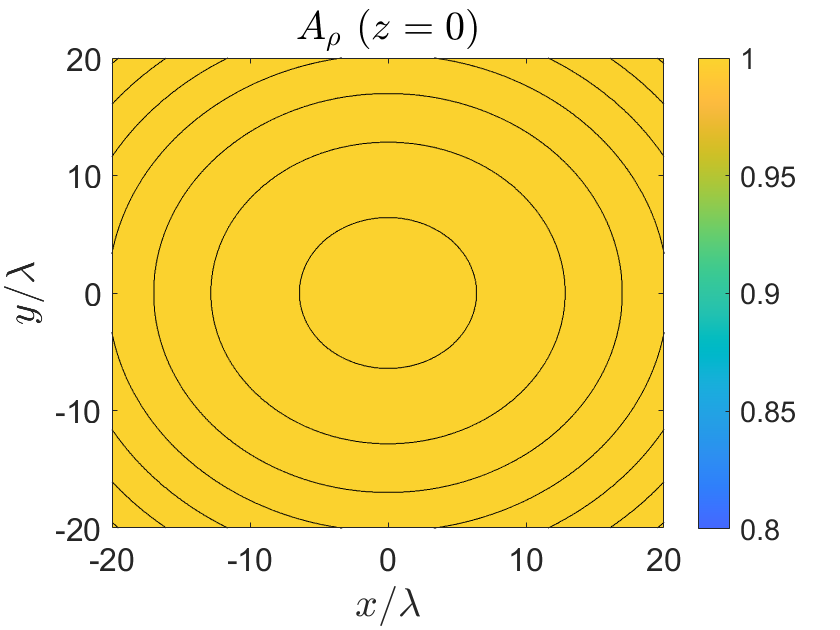}
\setcounter{subfigure}{1}}
\subfloat[]{\includegraphics[width=0.34\linewidth]{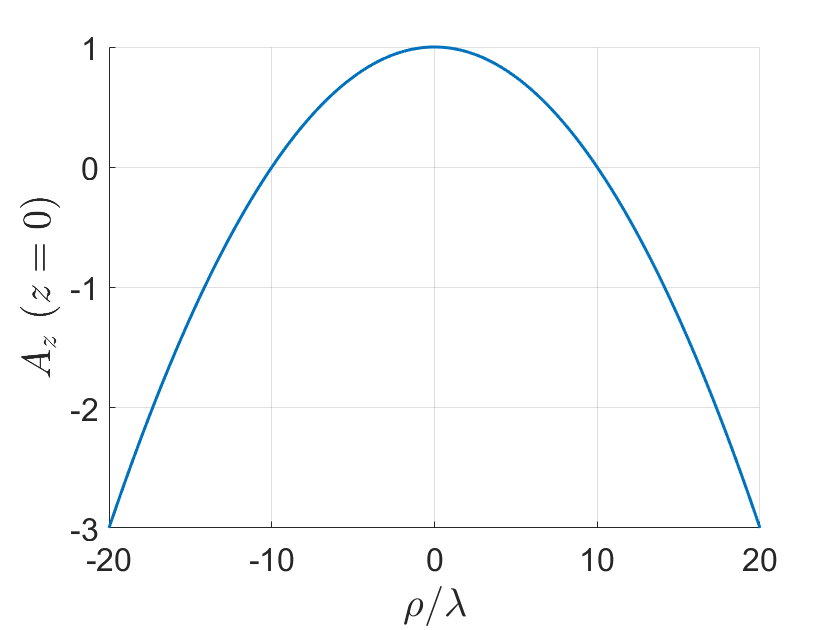}\setcounter{subfigure}{3}}
\subfloat[]{\includegraphics[width=0.34\linewidth]{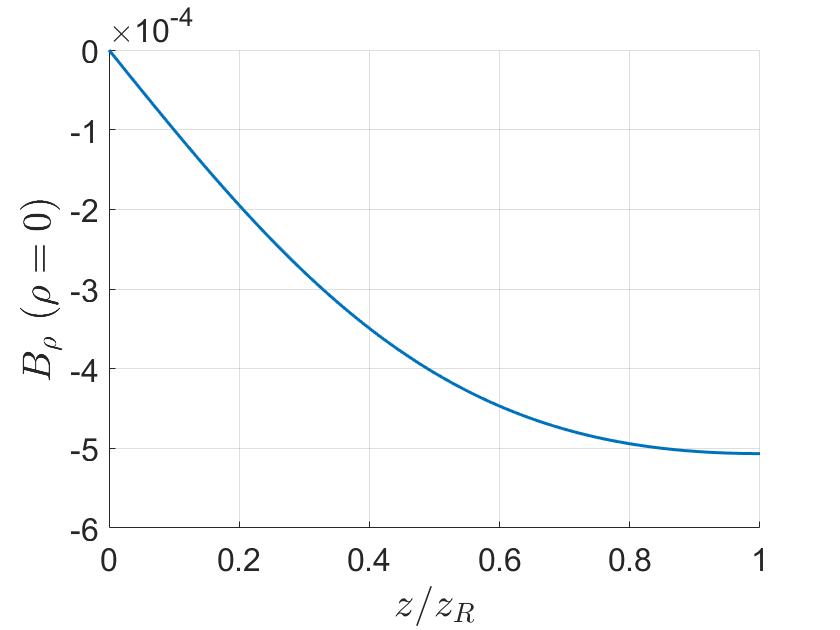}
\setcounter{subfigure}{5}}
\hfill
\subfloat[]{\includegraphics[width=0.34\linewidth]{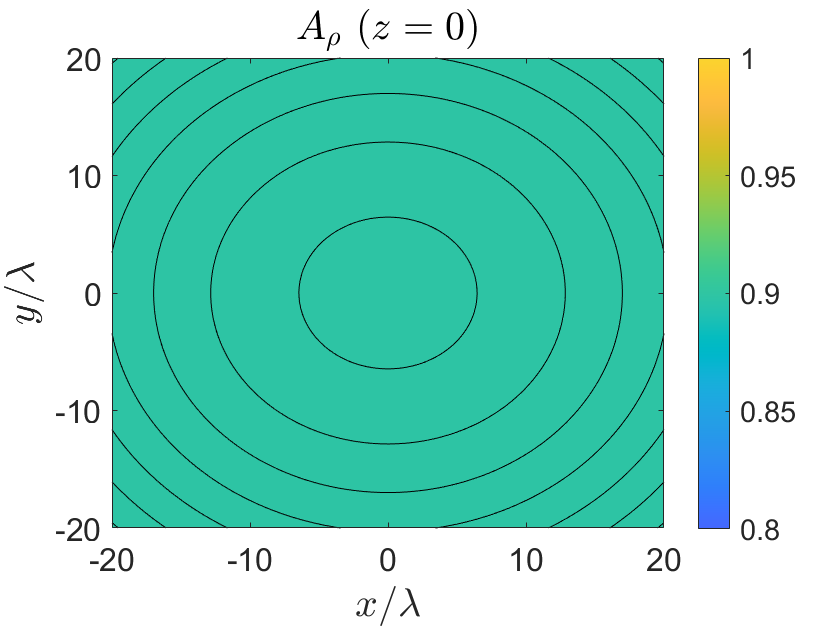}
\setcounter{subfigure}{2}}
\subfloat[]{\includegraphics[width=0.34\linewidth]{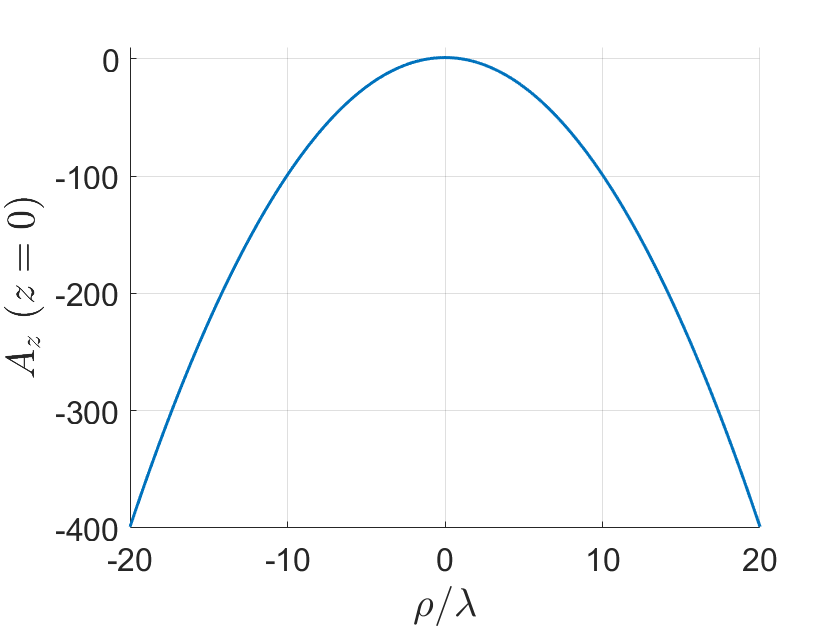}\setcounter{subfigure}{4}}
\subfloat[]{\includegraphics[width=0.34\linewidth]{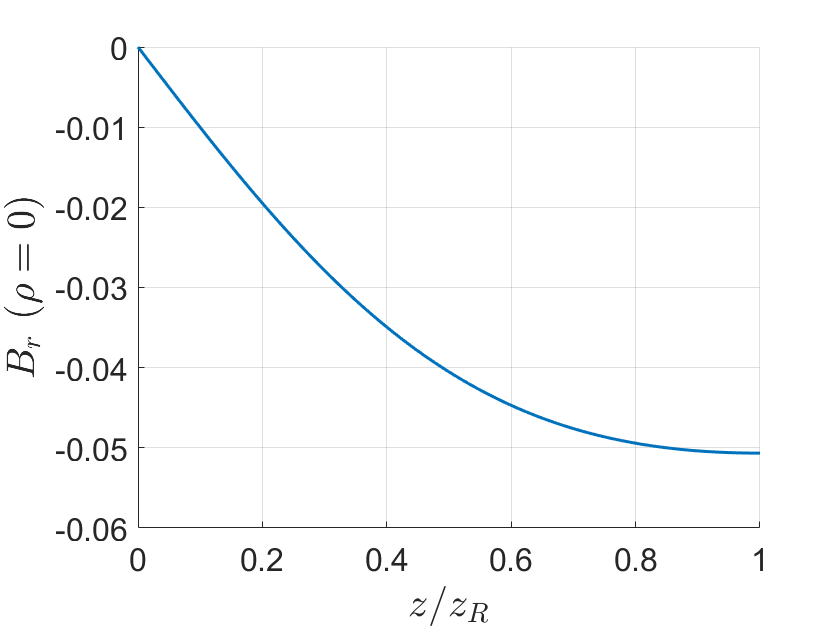}
\setcounter{subfigure}{6}}
 \caption{Depictions of some of the parameters from Eq.~(\ref{eq:SuppSimplification}) for a collimated beam (with $w_0=10\lambda$, upper row) and a focused beam (with $w_0=\lambda$, lower row), where $\lambda=400$ nm. Specifically, (a) and (b) depict $A_\rho$ in the transverse plane at the beam focus for $w_0=10\lambda$ and $w_0=\lambda$, respectively. $A_\rho$ is well approximated by unity at the beam focus in the former case and is approximately $0.9$ across the plane in the latter. (c) and (d) show the dependence of the term $A_z$ on $\rho/w_0$ at the beam focal plane $z=0$. It exhibits a shifted parabolic behavior whose $\rho$ dependence scales proportionally to $1/w_0^2$ as shown in Eq.~(\ref{eq:SuppSimplification}). The term $A_z\approx 1$ close to the beam axis in the case of a collimated beam but not for a focused beam (with $w_0=\lambda$). Finally, (e) and (f) display $B_\rho$ versus $z/z_R$ along the beam axis for the collimated and the focused beams, respectively. This term is very small for the collimated beam with $w_0=10\lambda$ and can be neglected when compared to $A_\rho$ and $A_z$, while this approximation is not good for beams with $w_0=\lambda$. As shown in Eq.~(\ref{eq:SuppSimplification}), $B_\rho$ scales as $1/(w_0\lambda)^2$. The parameter $B_\rho$ vanishes at the beam focus $z=0$.}
    \label{fig:Simplification}
\end{figure}

In this appendix, we show that for collimated beams, which have  $w_0 >> \lambda$, the real dimensionless parameters

\begin{equation}
\begin{array}{c}
    A_\rho = 1+\frac{1}{kz_{R}}\frac{\rho^{2}-2w_{0}^{2}}{w^{2}} +\left(\frac{2z\rho}{w^2 k z_R }\right)^2, \\
    B_\rho = -\frac{4}{w^{2}}\frac{1}{k^{2}}\frac{z}{z_{R}}\left(1-\frac{\rho^{2}}{w^{2}}\right), \\
    A_z = 1-\frac{\rho^2}{w^2}, \\
    B_z = \frac{z}{z_R}\frac{\rho^2}{w^2},
\end{array}
\label{eq:SuppSimplification}
\end{equation}

are approximately $A_\rho \approx 1$, $B_\rho \approx 0$, $A_z\approx 1$ and $B_z \approx 0$ close to the focus ($|z| \leq z_R$) and axis ($\rho^2 < w_0^2$) of the beam. 
This approximation simplifies the manipulation of collimated APBs, RPBs, ARPBs, and of their respective field quantities. The following simulations compare the values of $A_\rho$, $A_z$, and $B_\rho$ for a collimated APB with $w_0=10\lambda$ and a focused APB with $w_0=\lambda$. Figure~\ref{fig:Simplification}(a) and (b) show that $A_\rho \approx 1$ on the focus plane for the case of the collimated beam, and that $A_\rho \approx 0.9$ for the focused beam. Figure~\ref{fig:Simplification}(c) and (d) depict $A_z$ at the focal plane, with $A_z \approx 1$ for $\rho < w_0$ for the case of the collimated beam, which is not a good approximation for beams with $w_0=\lambda$. Fig.~\ref{fig:Simplification}(e) and (f) show that $B_\rho \approx 0$ on the beam axis is a good approximation for collimated beams, but less for focused beams. The value of $A_z$ on the beam axis and $B_z$ at the focus or axis of the beam are not shown as they are $1$ and $0$ respectively.

\ifCLASSOPTIONcaptionsoff
  \newpage
\fi

\bibliographystyle{IEEEtran}
\bibliography{IEEEabrv,references}